\documentclass[aps,prb,twocolumn,superscriptaddress]{revtex4-1}

\usepackage{bm,wrapfig}

\usepackage{graphicx}
\usepackage{epsfig}
\usepackage{amsfonts}
\usepackage{amsmath}
\usepackage{amssymb}
\usepackage{color}
\usepackage[colorlinks,linkcolor=blue,hyperindex,CJKbookmarks]{hyperref}
\usepackage{xcolor}

\newcommand\ba{\begin{eqnarray}}
\newcommand\ea{\end{eqnarray}}
\newcommand\be{\begin{equation}}
\newcommand\ee{\end{equation}}

\begin{document}
\title{Distinct signatures of particle-hole symmetry breaking in transport coefficients for generic multi-Weyl semimetals}

\author{Tanay Nag}
\email{tnag@physik.rwth-aachen.de}
\affiliation{Institut f\"ur Theorie der Statistischen Physik, RWTH Aachen University, 52056 Aachen, Germany}

\author{Dante M. Kennes}
\email{dkennes@physik.rwth-aachen.de}
\affiliation{Institut f\"ur Theorie der Statistischen Physik, RWTH Aachen University, 52056 Aachen, Germany}
\affiliation{Max Planck Institute for the Structure and Dynamics of Matter, Center for Free Electron Laser Science, 22761 Hamburg, Germany}

\begin{abstract}

We propose {and study} generic multi-Weyl semimetal (mWSM) lattice Hamiltonians that 
break particle-hole symmetry. 
{These models fall into two categories}: model I (model II) where the gap and tilt terms are coupled  (decoupled) can host
type-I and type-II Weyl nodes simultaneously (separately) in a hybrid  phase (type-I and type-II phases, respectively). We concentrate on the question of how anisotropy and non-linearity in the dispersions, gaps and tilt terms influence diffusive second order transport quantities namely, the circular photogalvanic effect (CPGE) and  the Berry curvature dipole (BCD) 
as well as first order Magnus Hall effect (MHE) in the  ballistic limit. The signatures of topological charges are clearly imprinted in the quantized CPGE response  for the hybrid mWSM phase in model I. Such a quantization is also found in the type-I WSM phase for model II, however, the frequency profiles of the CPGE in these two cases is distinctively different owing to their different band dispersion irrespective of the identical topological properties.
The contributions  from the vicinity of Weyl nodes and away from the WNs are clearly manifested in the BCD response, respectively, for model I model II.  The  Fermi surface properties for the activated momentum {lead to a few hallmark features} on the MHE  for
both the models. Furthermore, 
we identify distinguishing  signatures of the above responses for type-I, type-II and hybrid phases to provide an experimentally viable probe to  differentiate these WSMs phases.  

\end{abstract}

\maketitle

\section{Introduction}

Recent years have witnessed  a surge of studies of topological systems such as, topological insulator \cite{Hasan10} and topological superconductor \cite{sczhang11}, as they exhibit
exotic gapless edge states while the bulk remains gapped. In addition,  gapless bulk 
modes are noticed for  Weyl semimetals (WSMs)\cite{Burkov11,Wan11,yan2017topological,Armitage18}, Dirac semimetals, and nodal line semimetals \cite{Leon_11} hosting topologically protected surface states.
Interestingly, either by breaking time reversal symmetry (TRS) or inversion symmetry (IS), each twofold degenerate Dirac cone in Dirac semimetals {is split}
into two isolated gap closing points known as, Weyl nodes (WNs) of opposite chiralities 
\cite{McCormick17}. 
In particular, one can find at least two WNs of opposite
chirality when the system breaks TRS while this minimum number becomes four if the system breaks IS only.
These WNs, protected by some crystalline symmetries, carry a topological charge $n$ (quantified by the absolute value of the Chern number $ |{\mathcal C}|$) that is a quantized Berry flux through the Fermi surface enclosing it in the Brillouin zone (BZ) \cite{Armitage18}. The WSMs can be further classified into type-I and type-II WSMs: 
point-like (non point-like) Fermi surface at the WNs refers to type-I
(type-II) WSMs \cite{VOLOVIK2014514,YXu15,Soluyanov2015} {(however, note that one can define additional classes}  \cite{xpli21,sims2021topologically}). The
large tilting in the conical spectrum 
of the Weyl cone results in a Lifshitz transition from a type-I to a type-II class WSM where Lorentz invariance is no longer satisfied. 
In experiments, several inversion asymmetric compounds,  such as TaAs (for the type-I WSM) and  MoTe$_2$, WTe$_2$ (for the type-II WSMs)
have been synthesized \cite{lv2015observation, xu2015discovery, jiang2017signature, li2017evidence, kimura2019optical}.


Surprisingly, in contrast to the conventional WSMs with $n=1$, it
has  recently been shown, using first principles calculations,  that $n$ can be generically greater than unity \cite{Gxu11,bernevig12,Yang2014,Huang1180}. These are referred to as the multi WSMs (mWSMs) where 
the quasi-particle
dispersion becomes anisotropic and non-linear similar to the dispersion in multi-layer stacked graphene \cite{McCann06,Min08}. For example,  HgCr$_2$Se$_4$ and SrSi$_2$ are candidate materials for double WSM with $n=2$ \cite{bernevig12} while Rb(MoTe)$_3$ might be a triple WSM with $n=3$ \cite{liu17}. 
However, the experimental discovery of mWSMs is yet to be made. Remarkably, another class of WSM is given by the case where one WN belongs to the type-I while its chiral partner belongs to the type-II class. This kind of hybrid WSM, consisting of  mixed types
of WNs, has been theoretically predicted for single WSM \cite{Soluyanov2015,YXu15,fyli16}. In order to obtain the above phase, one requires to break particle-hole (PH) symmetry for  the individual WNs. 
To the best of our knowledge it remains an open question how to formulate a mWSM lattice model that hosts the above introduced hybrid mWSM phase in the presence of a PH symmetry breaking term. 

In the field of transport phenomena 
in topological systems, WSMs have also emerged as a fertile ground for theoretical   \cite{karl14,Sharma16} as well as experimental  \cite{cwang16,hirschberger2016chiral,Watzman18} research. The Fermi arc surface states, connecting the 
two WNs with opposite chiralities, {are responsible} for the topological transport properties \cite{Wan11}. To name a few, negative
magnetoresistance related to the chiral-anomaly, and the quantum anomalous Hall
effect \cite{Gxu11,Zyuzin12,Son_2013,Burkov_2015} are identified. In addition to electronic transport phenomena, there exist a plethora of studies on the thermal transport properties  
\cite{landsteiner2014anomalous,rlundgren14, sharma2017nernst,Nag_2020, zhang2018strong, hirschberger2016chiral, watzman2018dirac}. Furthermore,
The  electronic and thermal transport properties of type-II WSMs can be significantly different compared to type-I WSMs \cite{fei2017nontrivial, yu2016predicted,wang2016gate, nandy2017chiral,Sun_2018,Nag20, schindler2020anisotropic,gsharma17}. On the other hand, the anisotropic nature of non-linear dispersion can further alter  the transport properties as observed for
mWSMs \cite{Chen16,xlio16,evgorbar17,Dantas2018,jrwang19,nag2020magneto,das21}. {The realm of diffusive transport phenomena is further enriched by the following second order responses apart from the above mentioned first order transport coefficients.} The circular photogalvanic effect 
(CPGE) \cite{de2017quantized,konig17,flicker18,ni2020linear,ni2021giant,sadhukhan21,sadhukhan21b,Matsyshyn21} and Berry curvature dipole  (BCD) \cite{sodemann15,xu2018electrically,rostami18,facio18,zeng20,Zeng21,roy2021non,Matsyshyn19} mediated optical and  electronic effects, respectively,  emerge due to their unique response characteristics. To add even more, the concept of third order response was introduced very recently  \cite{hliu21,lai2021third}. Interestingly, some of these non-linear effects are  found to survive even when the TRS is not broken explicitly unlike the first order responses. 
Another first order transport mechanism in the presence of built-in electric field, namely the Magnus Hall effect (MHE), falls into the same category \cite{papaj19,Mandal20,das21,kapri2021magnus} {where the  electron transport is attributed to the Magnus velocity in absence of magnetic field. Note that MHE is a ballistic transport phenomena.}


Here, we aim at addressing the 
distinct signatures of the second order optical (i.e., CPGE)  and electrical (i.e., BCD) transport properties as well as of the first order (i.e., MHE) responses in mWSM concentrating on how  the anisotropy, non-linearity and tilt of the dispersion affect these signatures compared to the single WSM cases. The  topological charge is clearly imprinted on the magneto-transport behavior \cite{Dantas2018,nag2020magneto}. { One can distinguish TRS broken single WSMs from the TRS invariant counterpart by investing the CPGE  \cite{sadhukhan21,sadhukhan21b}}. Therefore, it is worth  studying the transport properties in TRS broken mWSM including a PH symmetry breaking tilt term which to the best of our knowledge remains an open question. 
To be more concrete, we answer the following questions: Can we distinguish the CPGE responses in the hybrid phase from that in the type-I and type-II  phases of mWSMs under suitably breaking the PH symmetry? How do the changes in Fermi surface characteristics,  influenced by such PH symmetry breaking terms, manifest in BCD and MHE responses?

In this work, we propose a generic mWSM lattice Hamiltonian, dubbed model I, where 
the  hybrid phase in addition to the type-I and type-II phases can be realized by appropriately tuning the PH symmetry breaking tilt term (Eqs.~(\ref{single_WSM}), (\ref{double_WSM}), and (\ref{triple_WSM})). We further investigate another model, referred to as model II, where the tilt and the gap terms are decoupled (Eq.~ (\ref{modelII_n0})) resulting in the non-degenerate WNs. This allows us  to compare different transport properties between these two models. We show that  the 
CPGE is found to be quantized, being proportional to the topological charge,  in the hybrid (type-I) phase for model I (model II) [see Figs.~\ref{fig:CPG_model1} and \ref{fig:comparison_CPG_typeII}]. The choice  of gap and tilt terms in both of the models allows us to explore the rich frequency profile of the CPGE as the effects of these terms are encapsulated in the Fermi distribution function  as well as the 
optically activated momentum surfaces. 
Turning our attention to another second order response namely, BCD, we find that  the mirror symmetry restricted diagonal components obtain significant contributions from the vicinity of WNs and  away from the WNs for model I and model II, respectively (see Figs.~\ref{fig:bcd_modelI} and \ref{fig:bcd_modelII}). These responses grow with increasing topological charge as the corresponding Fermi surface contribution enhances. We find that the {Magnus Hall conductivity (MHC),  connected to the  MHE,} also noticeably changes between model I and II, depending on the nature of the tilt and gap term, as the   
the distribution of ballistically activated momentum modes and the associated  Fermi surface  profiles are modified (see Fig.~\ref{fig:mhc_modelI} and \ref{fig:mhc_modelII}). We also thoroughly distinguish the type-II response from that for type-I to provide guidance to experiments on how to distinguish these phases.

The paper is organized as follows. In Sec.~\ref{sec2}, we describe the generic models for tilted mWSM namely, model I and model II. Next in  Sec.~\ref{CPG_formalism}, we discuss the formalism to compute the second order response CPGE along with our findings. We then illustrate the BCD induced second order response properties for our 
models in Sec.~\ref{bcd}. After that, we analyze the MHC in Sec.~\ref{MHC}. We compare our results with the existing literature in Sec.~\ref{discussion}. We extend  the discussion on material and experimental  connections in Sec.~\ref{material} and \ref{experiment}, respectively.
Finally, in Sec.~\ref{conclusion}, we conclude with possible future direction.

\section{Lattice Hamiltonian for  mWSM}

\label{sec2}

\subsection{Model I: Hybrid mWSM}
\label{lattice_model1}

We start with a two-band tight-binding model  on a cubic lattice.
The general form of the Hamiltonian in  momentum space can be written as follows: 
\begin{equation}
{\mathcal H}({\bm k})=\mathbf{N}_{{\bm k}}\cdot \boldsymbol{\sigma}
\end{equation}
with $\boldsymbol{N}_{{\bm k}}=[N_x,N_y,N_z]$ and pseudo-spin
$ \boldsymbol{\sigma}=[\sigma_x,\sigma_y,\sigma_z]$. We consider TRS breaking $\mathcal{T} {\mathcal H}({\bm k}) \mathcal{T}^{-1} \ne {\mathcal H}(-{\bm k}) $ with $\mathcal{T}= \mathcal{K}$ such that the lattice model hosts two degenerate WNs. Here, $\mathcal{K}$ denotes the complex conjugation operation.  In order to break the degeneracy, one needs to incorporate $N_0 \sigma_0$ in ${\mathcal H}(\mathbf{k})$: 
$H({\bm k})= {\mathcal H}({\bm k}) + N_0 \sigma_0$. 
Our  aim is to add an appropriate $N_0$ such that the IS, generated by $\mathcal{P} =\sigma_z$, and anti-unitary PH symmetry, generated by $\mathcal{A} = \sigma_x \mathcal{K}$, are broken: 
$\mathcal{P} H( {\bm k}) \mathcal {P}^{-1}  \ne H (-{\bm k}) $  and  $\mathcal{A} H( {\bm k}) \mathcal {A}^{-1}  \ne -H(-{\bm k}) $. These symmetry breakings will determine the nature of the phases in WSMs \cite{fyli16}. For a certain phase, two WNs can show different tilt configuration i.e., left WN can be of type-I while  right WN can be of type-II. Below we explicitly demonstrate the lattice models for single, double and triple WSMs where different phases can be found. 

\begin{figure} [t] 
\includegraphics[width =0.32\columnwidth]{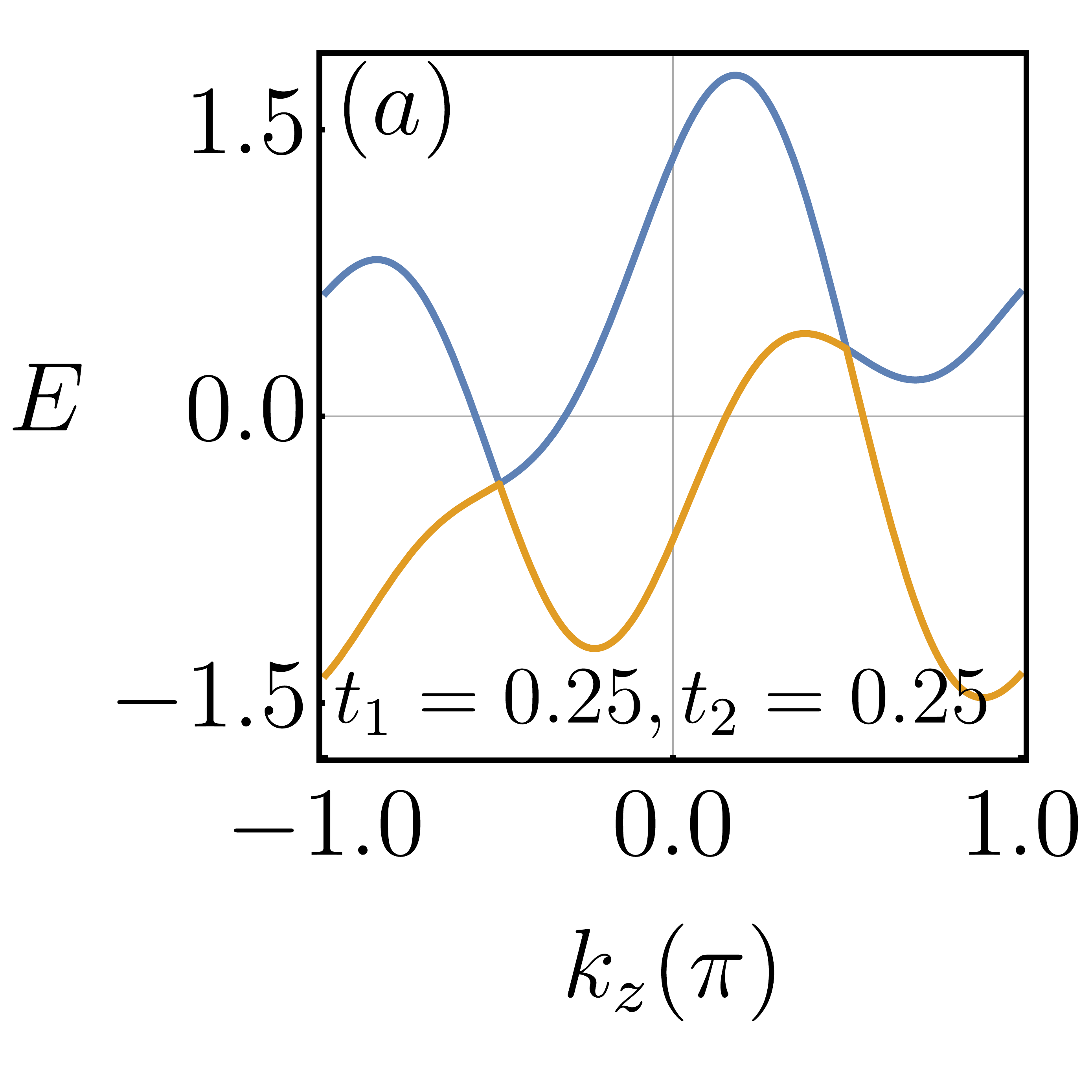} 
\includegraphics[width =0.32\columnwidth]{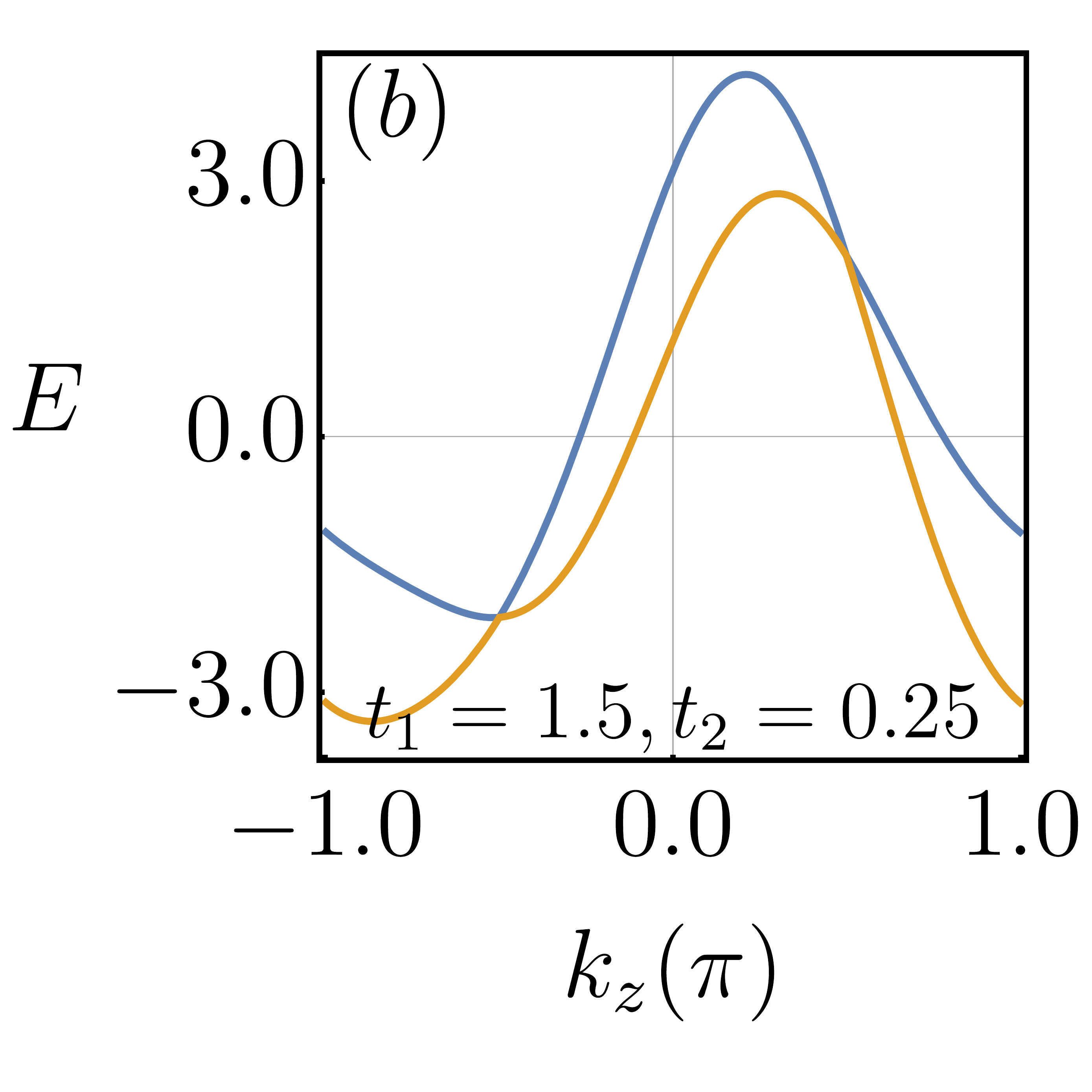} 
\includegraphics[width =0.32\columnwidth]{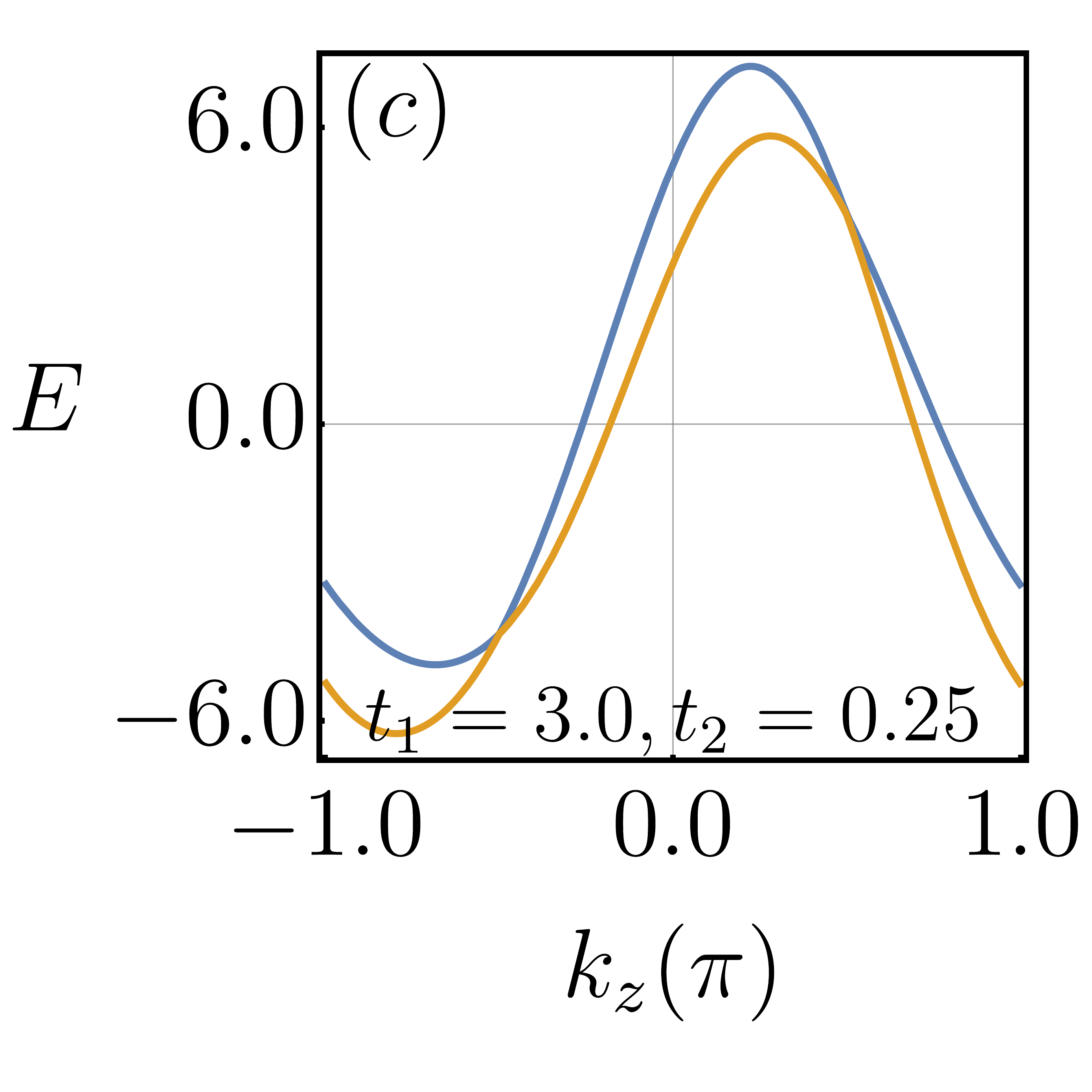} 
\caption{ The dispersions of single WSM, following Eq.~(\ref{single_WSM}),  are shown for $t_1=0.25$ in (a), for $t_1=1.5$ in (b) and  for $t_1=3.0$ in (c). 
(a) The  WNs at negative and positive energies are both of type-I referring to the type-I phase. (b) The negative (positive) energy WN is of type-I (type-II) corresponding to the hybrid phase. (c)
The WNs at negative and positive energies both are of type-II suggesting to the type-II phase. 
We consider $t_2=0.25$, $(\phi_1,\phi_2)=(\pi/4,\pi/2)$ and $k_x=k_y=0$. The dispersions for the double and triple WSM look the same as 
these dispersion  do not change along $k_z$ with increasing topological charge given the above set of parameters (not shown). The  WN at $k_z=\pi/2$ with positive energy is more tilted as compared to its counterpart at $k_z=-\pi/2$ for $t_{1,2} \ne 0 $. 
}
\label{fig:model-I}
\end{figure}

For the single-WSM with topological charge $n=|{\mathcal C}|=1$ described by Hamiltonian $H_{n=1}$, $\mathbf{N}_{\bm k}$ is chosen as 
\begin{eqnarray}
 N_0&=&2 t_1 \cos(\phi_1 - k_z) +  2 t_2 \cos(\phi_2 - 2 k_z), \nonumber \\
 N_x&=&t \sin k_x,~~ N_y=t \sin k_y, ~~~\rm{and}\nonumber \\
 N_z&=& t_z \cos k_z -m_z +t_0(2-\cos k_x -\cos k_y).
\label{single_WSM}
 \end{eqnarray}
Note that $N_0$ includes a first and second nearest neighbour pseudo-spin independent hopping along the $z$-direction denoted by $t_1 \exp(-i \phi_1)$ and $t_2 \exp(-i \phi_2)$, respectively \cite{fyli16}. We include a phase difference between these complex  hopping terms, allowing us to modulate  the energies as well as tilt of the WNs. Importantly, position and chirality of the WNs, determined by the $N_{x,y,z}$ terms, remain unaltered irrespective of the choice of $N_0$. In this model, the WNs are located at ${\bm k}=(0,0, sk_0)$ with 
\begin{equation}
\cos (s k_0)=\frac{t_0}{t_z}\bigl[ \frac{m_z}{t_0}+\cos k_x +\cos k_y -2\bigr]
\label{eq_lm1}
\end{equation}
and $s=\pm$.
One can expand the above Hamiltonian around $k_z=sk_0$ with $m_z=0$, $t=t_0=t_z=1$ to obtain the low energy 
Weyl Hamiltonian: $H_{n=1,s}\approx 2 k_z(t_1 \sin(\phi_1 - s k_0) + 2 t_2 \sin(\phi_2 - 2 s k_0) ) \sigma_0+ t( \sigma_x k_x +  \sigma_y k_y)+ s t_z  \sigma_z k_z \sin k_0$.  Importantly, from the low-energy model $H_{n=1,\pm }$, one directly finds  ${\mathcal C}=\mp 1$.   
The quantity $\eta = |2 (t_1 \sin(\phi_1 - s k_0) + 2 t_2 \sin(\phi_2 - 2s k_0) )/(s t_z  \sin k_0)|$, representing the tilt strength,
determines whether the single WSM resides in the type-I or type-II phase;  $\eta<1$ ($\eta>1$) corresponds to type-I (type-II) single WSM when both the WNs
for $s=\pm $ with ${\mathcal C}=\mp 1$  exhibit similar tilt profiles. 
Interestingly,   
the hybrid phase arises when the WN for $s=- $ with ${\mathcal C}=+1$
behave distinctly from the WN for $s=+ $ with  ${\mathcal C}=-1$. To be precise,  the  WN at $k_z=-\pi/2$ is of type-I (type-II) while the right WN at $k_z=+\pi/2$ belongs to type-II (type-I). This is clearly illustrated in Fig.~\ref{fig:model-I}. 
One can thus obtain type-I, type-II and hybrid phases by appropriately tuning the  parameters
$t_{1,2}$ and $\phi_{1,2}$.


Now turning towards the double WSM with topological charge $n=|{\mathcal C}|= 2$ described by Hamiltonian $H_{n=2}$, $\mathbf{N}_{\bm k}$ acquires the following form\cite{Roy_2017} 
\begin{eqnarray}
 N_0&=&2 t_1 \cos(\phi_1 - k_z) +  2 t_2 \cos(\phi_2 - 2 k_z), \nonumber \\
 N_x&=&t(\cos k_x-\cos k_y),~~ N_y=t\sin k_x \sin k_y, ~~~\rm{and} \nonumber \\
 N_z&=& t_z \cos k_z -m_z + t_0(6+\cos 2k_x
+\cos 2k_y \nonumber \\
&-& 4 \cos k_x -4 \cos k_y).
\label{double_WSM}
 \end{eqnarray}
The lattice model of double WSM contains two WNs at ${\bm k}=(0,0, sk_0)$, similar to the single WSM, with 
\begin{align}
 \cos (sk_0)=\frac{t_0}{t_z}\bigl[ \frac{m_z}{t_0}-(&6+\cos 2k_x +\cos 2k_y 
 \notag\\
 &-4\cos k_x -4\cos k_y)\bigr]
\label{eq_lm2}
\end{align}
One can similarly expand the above Hamiltonian around $k_z= sk_0$ and $m_z=0$, $t=t_0=t_z=1$ as stated for the single WSM. In this case, the low energy Hamiltonian for double-WSM
with a given  $s$
can be written as $H_{n=2,s}\approx 2 k_z(t_1 \sin(\phi_1 - s k_0) + 2 t_2 \sin(\phi_2 - 2 s k_0) ) \sigma_0+ \frac{t}{2}( \sigma_x (k_x^2-k_y^2) + \sigma_y k_x k_y))+ s t_z \sin k_0 \sigma_z k_z$. 
Similarly, for a triple-WSM with topological charge $n=|{\mathcal C}|=3$ described by Hamiltonian $H_{n=3}$, the $\mathbf{N}_{\bm k}$ is given by 
\cite{Roy_2017} 
\begin{eqnarray}
N_0&=&2 t_1 \cos(\phi_1 - k_z) +  2 t_2 \cos(\phi_2 - 2 k_z), \nonumber \\
 N_x&=&t\sin k_x(1-\cos k_x-3(1-\cos k_y)), \nonumber \\
 N_y&=&-t\sin k_y (1-\cos k_y -3(1-\cos k_x)), ~~~\rm{and} \nonumber \\
 N_z&=& t_z \cos k_z -m_z + t_0(6+\cos 2k_x
+\cos 2k_y \nonumber \\
&-& 4 \cos k_x -4 \cos k_y).
\label{triple_WSM}
 \end{eqnarray}
The low energy triple-WSM Hamiltonian is  given by
$H_{n=3,s}\approx  2 k_z(t_1 \sin(\phi_1 - s k_0) + 2 t_2 \sin(\phi_2 -2 s k_0) ) \sigma_0 + \frac{t}{2}( \sigma_x (k_x^3-3k_x k_y^2) - \sigma_y(k_y^3- 3k_x^2 k_y))+ s t_z \sin k_0 \sigma_z k_z$. 
Combining single, double and triple WSMs,
the general form of the low-energy Hamiltonian for a topological charge $n$ is  given by 
$H_{n,s,{\bm k}} = 2 k_z(t_1 \sin(\phi_1 - s k_0) + 2 t_2 \sin(\phi_2 - 2s k_0) ) \sigma_0 +  \alpha k_+^n \sigma_+ +\alpha k_-^n \sigma_- + s t_z k_z \sin k_0 \sigma_z $ with $k_{\pm} = k_x \pm ik_y$ and $\sigma_{\pm}=( \sigma_x \pm i \sigma_y)/2$. 
We note that the type-I, type-II and hybrid phases are observed for double and triple WSMs similar to the single WSM.


It is also  noted that single and triple WSMs without the term $N_0\sigma_0$ preserve the inversion and anti-unitary PH symmetry. The identity term can cause the WNs for $k_z=s k_0=s\pi/2$ to appear at  different WN energies $E_s= 2 s t_1 \sin \phi_1 - 2 t_2 \cos \phi_2$. The chirality, associated with the WN at $k_z=\pi/2$ ($k_z=-\pi/2$), is ${\mathcal C}=-1$, and $-3$ (${\mathcal C}=+1$, and $+3$), respectively, for single and triple WSM while for double WSM, the WN at $k_z=\pi/2$ ($k_z=-\pi/2$), corresponds to ${\mathcal C}=+2$ (${\mathcal C}=-2$). For sake of simplicity, we below demonstrate the effect of $N_0$ for the single WSM in more details. This discussion  can be carried over to mWSMs upon appropriately incorporating the chiralities of individual WNs.   Note that for $t_{1,2}>0$, the WN at $k_z=\pi/2$ ($k_z=-\pi/2$) has tilt strength 
$\eta=4 t_2 +\sqrt{2} t_1$ ($\eta=|4 t_2 -\sqrt{2} t_1|$). As a result, the WN with positive energy at $k_z=\pi/2$ is more tilted as compared to the other WN with negative energy at $k_z=-\pi/2$.  
This is clear from the  structure of $\eta$ that remains unaltered for single, double and triple WSMs. 

\begin{figure} [t] 
\includegraphics[width =0.8\columnwidth]{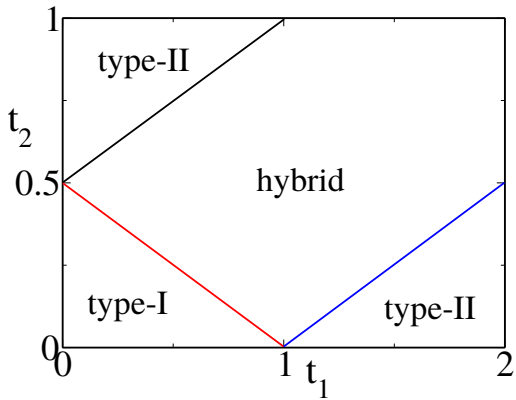} 
\caption{ The topological phase diagram of model I, considering the single WSM Eq.~(\ref{single_WSM}). The red, blue and black line correspond to the following equations $t_2 = 0.5 - 0.5  t_1$, $t_2 = -0.5 + 0.5  t_1$ and $t_2 = 0.5 +0.5  t_1$, respectively. We consider $(\phi_1,\phi_2)=(\pi,\pi/2)$. Note that this phase diagram remains unaltered for the double and triple WSMs cases. 
}
\label{fig:model-I_pd}
\end{figure}

Interestingly, for the choice of the following parameter $\phi_1=\pi$, $\phi_2 =\pi/2$, the WNs become degenerate and this degeneracy is not protected by any symmetry in the general situation. With the above set of parameters, one can evaluate  
$\eta = |2 (t_1 \sin(\phi_1 - s k_0) + 2 t_2 \sin(\phi_2 -2 s k_0) )/(s t_z  \sin k_0)|$  to 
determine the type-I and/or type-II nature  of 
WN associated with the chirality ${\mathcal C}=\pm 1$.
Considering $t_{1,2}>0$,
both the WNs at $k_z=\pm \pi/2$ are of type-I i.e., $\eta<1$ for ${\mathcal C}=\mp 1$,  when $t_2 <  -0.5  t_1 + 0.5$. The type-II phase i.e., $\eta >1$ for  both the WNs with ${\mathcal C}=\mp 1$ at $k_z=\pm \pi/2$
is separated by the phase boundaries $t_2 >  0.5 t_1 +0.5 $ and $t_2 < -0.5 + 0.5  t_1$. 
On the other hand, the hybrid phase is bounded by: $t_2 <  0.5  t_1 +0.5 $, $t_2 > -0.5 + 0.5  t_1$ and $t_2 >  -0.5 t_1 + 0.5$ within which WN at $k_z=-\pi/2$ ($k_z=\pi/2$) is of type-I i.e., $\eta<1$ with ${\mathcal C}=+1$ (type-II i.e., $\eta>1$ with ${\mathcal C}=-1$) \cite{fyli16}. The topological phase diagram, including type-I, type-II and hybrid WSM phases, is shown in Fig.~\ref{fig:model-I_pd}.

For a given set of parameters $(t_1,t_2)$,
the WNs  continue to appear at the same energy as long as $\phi_1=m \pi$ with $m=0,~1,~2, ~\cdots$, irrespective of the value of $\phi_2$. On the other hand, for  $\phi_2=(2m+1) \pi/2$, $\phi_1 \ne m\pi$, two WNs are always separated in energy. 
The maximum energy separation between two WNs appear when $\phi_1=(2m+1) \pi/2$ and   $\phi_2 = m\pi$.
We consider $(\phi_1,\phi_2)=(\pi/4,\pi/2)$ to tune the WNs to different  energies.  We use the following set of parameters for the model I: $t=t_z=1$, $m_z=0$, $t_2 =0.25$, $\phi_1=\pi/4$, and $\phi_2=\pi/2$ thorough out our paper. 
The type-I phase, irrespective of the topological charge $n=|{\mathcal C}|$,  is found for  $t_1 <0.71$, hybrid phase for $0.71< t_1 <2.12$, and type-II phase for $t_1 >2.12$ as shown in Fig.~\ref{fig:model-I}. In this case, the WNs with chirality ${\mathcal C}=\pm 1,\pm 3 $ appear at energies $E_{\mp}=\mp \sqrt{2} t_1$ at momentum $k_z=\mp \pi/2$ for single and triple WSMs. For double WSM,  WNs of chirality ${\mathcal C}=\pm 2 $ appear at energies $E_{\pm}=\pm \sqrt{2} t_1$ at momentum $k_z=\pm \pi/2$. One can thus easily understand that the hybrid phase in single and triple WSMs hosts type-I (type-II) WN at $k_z=-\pi/2$ ($k_z=\pi/2$) with ${\mathcal C}=+1$ and $+3$ (${\mathcal C}=-1$ and $-3$), respectively.  
On the other hand, in the hybrid phase of double WSM, WN at $k_z=\pi/2$ ($k_z=-\pi/2$) is of type-I i.e., $\eta<1$ with ${\mathcal C}=+2$ (type-II i.e., $\eta>1$ with ${\mathcal C}=-2$).

\subsection{Model II: Conventional mWSM}
\label{lattice_model2}

\begin{figure} [t] 
\includegraphics[width =0.49\columnwidth]{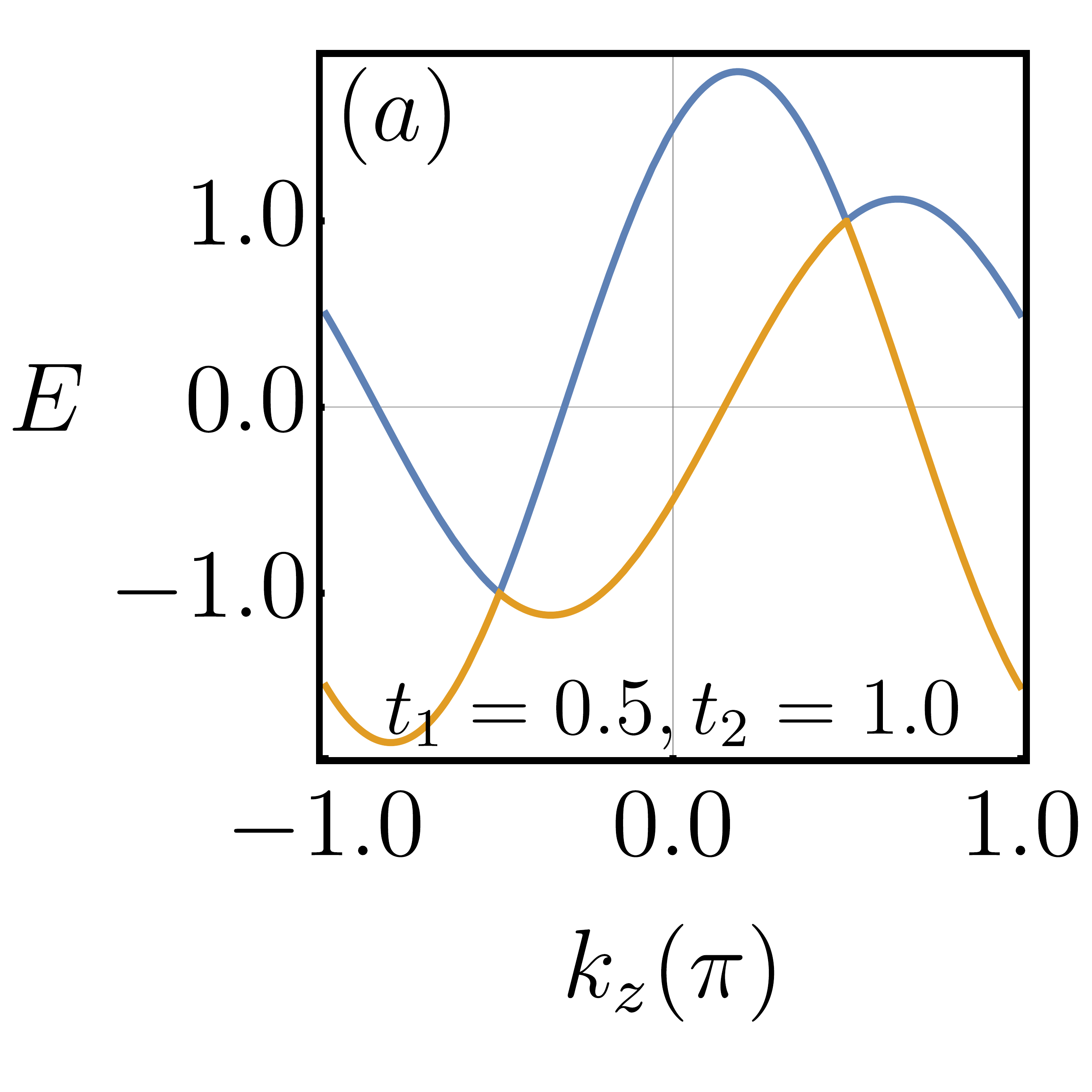} 
\includegraphics[width =0.49\columnwidth]{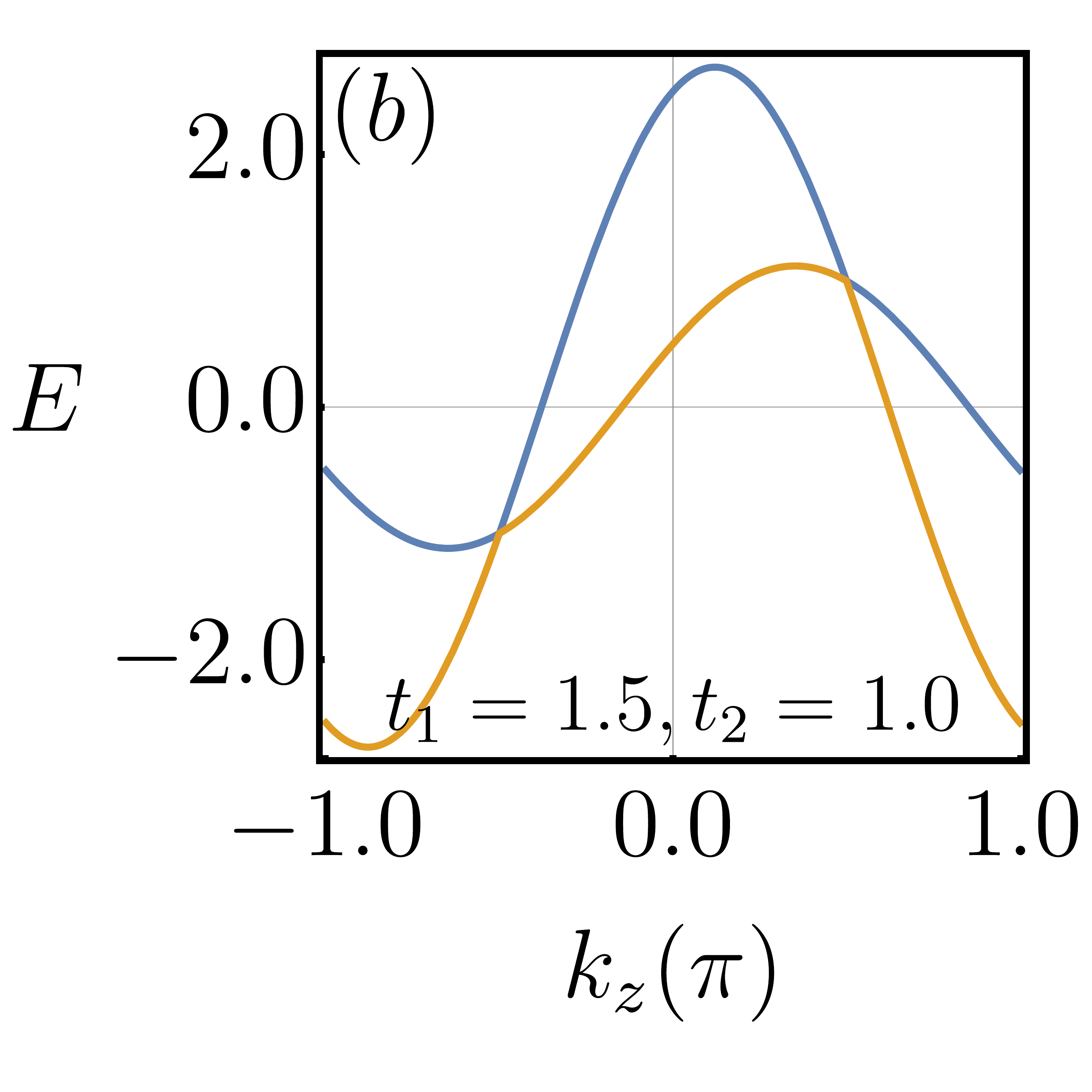} 
\caption{ The dispersions of single WSM, following Eq.~(\ref{single_WSM}) with $N_0$ given in Eq.~(\ref{modelII_n0}),  are shown for type-I phase with $t_1=0.5$ in (a) and type-II phase with $t_1=1.5$ in (b). Unlike the model I, the hybrid phase does not exist in model II. We consider $t_2=1$ and the remaining parameters are same as Fig.~\ref{fig:model-I} such that $k_0=\pi/2$.
The non-degenrate WNs appear at $E_{\pm}=\pm t_2$.
Unlike the dispersion in model I, shown in Fig.~\ref{fig:model-I},  the WNs with positive and negative energies at $k=\pi/2$ and $k=\pi/2$ both exhibit    identical tilt strength for model II. 
}
\label{fig:model-II}
\end{figure}

Having described the hybrid phase in model I, we here consider another form of $N_0$ so that gap and tilt terms are decoupled calling this model II:
\begin{equation}
N_0= t_1 \cos k_z + t_2 \sin k_z.
 \label{modelII_n0}
\end{equation}
Here the gap is given by the hopping $t_2$ that causes the non-degenerate WNs. The tilt is controlled by $t_1$ only.
The single, double and triple WSM lattice Hamiltonians for model II are same as given in Eqs.~(\ref{single_WSM}), (\ref{double_WSM}), (\ref{triple_WSM}) with
$N_0= t_1 \cos k_z + t_2 \sin k_z$.
We note that the IS and PH symmetry are both broken by the   above term.  The type-I [type-II] phase is observed for $\eta=
|(t_2 \cos k_0  - s t_1 \sin k_0 )
/(s t_z  \sin k_0)| <1$ [$\eta=|(t_2 \cos k_0  - s t_1 \sin k_0 )/(s t_z  \sin k_0)| >1$] while the WNs appear at $k_z=s k_0$ with $s=\pm$. The hybrid phase no longer exists here as both the WNs of opposite chiralities share an identical tilt profile.
This is intimately related to the fact that the tilt term $t_1$ eventually becomes 
decoupled from  $s$  while the gap term is still connected with $s$ as evident from $\eta$. 
This is in contrast to the model I where tilt and gap terms are mutually coupled with $s$. 
The type-I and type-II phases are shown in Fig.~\ref{fig:model-II} where we consider $t_2=1$ such that the WNs are non-degenerate with energy $E_s= s \sin k_0 +t_1 \cos k_0$. The chiralities, associated with the individual WNs, for the single, double and triple WSMs in model II are identical to that in model I. We reiterate that in model II similar to model I, we consider $t=t_z=1$, $m_z=0$, such that $k_0=\pi/2$. Note that both the WNs at $k=\pm \pi/2$ exhibit identical tilt strength $\eta=t_1$ in model II unlike the model I.


The low-energy dispersion of a WN in model I and II with a given $s$ are, respectively, given by 
\begin{eqnarray}
\epsilon_{{\bm k},s}^{\pm} &=& 2k_z (t_1 \sin(\phi_1 - s k_0) + 2 t_2 \sin(\phi_2 -2 s k_0) ) \nonumber \\
&\pm& \sqrt{ \alpha^2 k^{2 n}_{\bot} + t_z^2 k^2_z \sin^2 k_0}    
\label{eq_vel_model1}
\end{eqnarray}
and 
\begin{eqnarray}
\epsilon_{{\bm k},s}^{\pm}&=&k_z(t_2 \cos k_0 - s t_1 \sin k_0)  \pm \sqrt{ \alpha^2 k^{2 n}_{\bot} + t_z^2 k^2_z \sin^2 k_0} \nonumber \\
\label{eq_vel_model2}
\end{eqnarray}
with $k_{\bot}=\sqrt{k^2_x+k^2_y}$.
The anisotropy in the dispersion is clearly visible as compared to the dispersion in untilted  single WSM  $ \epsilon_{{\bm k}}^{\pm}=\pm \sqrt{k_x^2+k_y^2+k_z^2}$. The quasi-velocity ($v_{\bm k}=\frac{\partial \epsilon_{\bm k}}{\partial {\bm k}}$)  is given by 
\begin{equation}
v_{\bm k}=\frac{1}{\epsilon_{{\bm k},n} }(k_x n \alpha^2 k_{\bot}^{2(n-1)},k_y n \alpha^2 k_{\bot}^{2(n-1)},k_z t_z^2 \sin^2 k_0 +\epsilon_{{\bm k},n} \gamma).
\label{eq_vel}
\end{equation}
where $\epsilon_{{\bm k},n}=\sqrt{ \alpha^2 k^{2 n}_{\bot} + t_z^2 k^2_z \sin^2 k_0}$. Here, 
$\gamma=2 (t_1 \sin(\phi_1 - s k_0) + 2 t_2 \sin(\phi_2 -2 s k_0) )$ and $\gamma=(t_2 \cos k_0 - s t_1 \sin k_0)$ for model I and II, respectively. The velocity for a single untilted WSM is
$v_{\bm k}={\bm k}/\epsilon^{\pm}_{\bm k}$ reflecting  the isotropic nature of the velocity in all momentum directions. Importantly, the $z$-component of velocity is different in these two models which can potentially lead to distinct response properties.

The Berry curvature (BC) of the m$^{\textrm{th}}$ band for a Bloch Hamiltonian $H({\bm k})$,
defined as the Berry phase per
unit area in the momentum ${\bm k}$ space, is given by ~\cite{PhysRevB.74.085308}
\begin{equation}
\Omega^{m}_{ {\bm k},a}= (-1)^m \frac{1}{4|N_{\bm k}|^3} \epsilon_{a b c} N_{\bm k} 
\cdot \left( \frac{\partial N_{\bm k}}{\partial k_b} \times \frac{\partial N_{\bm k}}{\partial k_c} \right) .
\label{bc}
\end{equation}
Interestingly, the $N_0$ term does not appear in the BC owed to the fact that the topological charges of these different models are identical. 
One can estimate the BC, following the low-energy models, as follows
\begin{equation}
{\Omega}_{\bm k}^{\pm} =\pm \frac{1}{2} \frac{n t_z\sin k_0 \alpha^2 k^{2n-2}_{\bot} }{|\epsilon_{{\bm k},n}|^{3}}
\: \left( k_x, k_y, n k_z \right).
\label{eq_bcl}
\end{equation}
This is markedly different from the 
BC of a single WSM ${\Omega}_{\bm  k}^{\pm}= \pm {\bm k}/|\epsilon^{\pm}_{\bm k}|^3 $. In particular, one notices that $\Omega_z$ depends  on $n$ in a quadratic manner while $\Omega_x$ and $\Omega_y$ are linearly dependent on  $n$. The Chern number ${\mathcal C}_m$, associated with the band index $m$, can be found by a closed surface  momentum integration of the BC:
\begin{equation}
 {\mathcal C}_m=\frac{1}{2 \pi}\int  {\Omega}^m_{\bm k} d^2{\bm k}.
\end{equation}
We  use Chern number as  chirality  interchangeably throughout the paper. ${\mathcal C}_{\pm}=\pm 1$, $\pm 2$ and $\pm 3$ in a single, double and triple WSM for valence ($-$) and conduction ($+$) band \cite{Dantas2018}.

\section{Circular photogalvanic effect (CPGE)}
\label{CPG_formalism}

The circularly polarized light induced second order optical response namely, the CPGE injection current is defined as
\begin{equation}
\dfrac{d J_i}{dt} = \beta_{ij}(\omega) \left[\mathbf{E}(\omega)\times \mathbf{E}^{*}(\omega)\right]_{j},
\label{injection}
\end{equation}
where $\mathbf{E}(\omega)=\mathbf{E}^{*}(-\omega)$ is the circularly polarized electric field of frequency $\omega$, $i$ and $j$ indices are the directions of current $J_i$ and the circular polarized light fields, respectively.  With the reversal in the polarization of the incident light i.e., exchange of $\mathbf{E}(\omega)$ and $\mathbf{E}^{*}(\omega)$, the 
photocurrent changes its sign.  We note that the photocurrent $J_i$ is a measure of the non-local diffusion of photo-excited carriers. 
The photocurrent is represented by the  CPGE tensor multiplied by carrier lifetime $\tau$. The CPGE tensor $\beta$ can be generically expressed as \cite{de2017quantized, sipe2000second} :
\begin{eqnarray}
\label{eq:nu}
\beta_{ij}(\omega) &=& \dfrac{\pi e^3}{\hbar V}\epsilon_{jkl}  \sum_{\bm k,n,m}  \Delta f_{\bm k,nm} {\Delta v} ^i_{\bm k,nm} r^k_{\bm k,nm} r^l_{\bm k,mn} \nonumber \\
&\times& \delta(\hbar\omega - E_{\bm k,mn}),
\end{eqnarray}
where $V$ is the sample volume, $E_{\bm k,nm}=E_{\bm k,n}-E_{\bm k,m}$ and $\Delta f_{\bm k,nm}=f_{\bm k,n}-f_{\bm k,m}$ are the difference between $n$-th and $m$-th band energies and Fermi-Dirac distributions respectively, 
$\mathbf{r}_{\bm k,nm} = i \left<n|\partial_{\bm k}|m\right>$ is the off-diagonal Berry connection and ${\Delta v}^i_{\bm k,nm} = \partial_{k_i}E_{\bm k,nm}/\hbar= v^i_{{\bm k},n}- v^i_{{\bm k},m}$. The trace of the CPGE tensor $\beta_{ij}$ for a two band model with $n,m=1,\;2$ is given by  
\begin{eqnarray}
{\rm Tr}[ \beta(\omega)] 
= \dfrac{ i \pi e^3}{\hbar^2 V} \sum_{{\bm k},i} \Delta f_{\bm k,12}  \Delta v^i_{{\bm k},12}\Omega_{{\bm k},i} \delta(\hbar\omega - E_{\bm k,12})\nonumber \\
\label{beta}
\end{eqnarray}
Here, $\Omega_{{\bm k},i}= i \epsilon_{ikl} \sum_{n\ne m} r^k_{\bm k,nm} r^l_{\bm k,mn} $ is the $i$-th component of the BC. We note that ${\rm Tr}[ \beta(\omega)]$ reverses its sign under reversal of polarization $k (l) \to l (k)$ due to the underlying anti-commutator like form of  $\Omega_{{\bm k},i}$ \cite{sadhukhan21b}. In our numerical analysis, we consider $\hbar=1$ without loss of generality.

\begin{figure*} [ht] 
\includegraphics[width = \textwidth]
{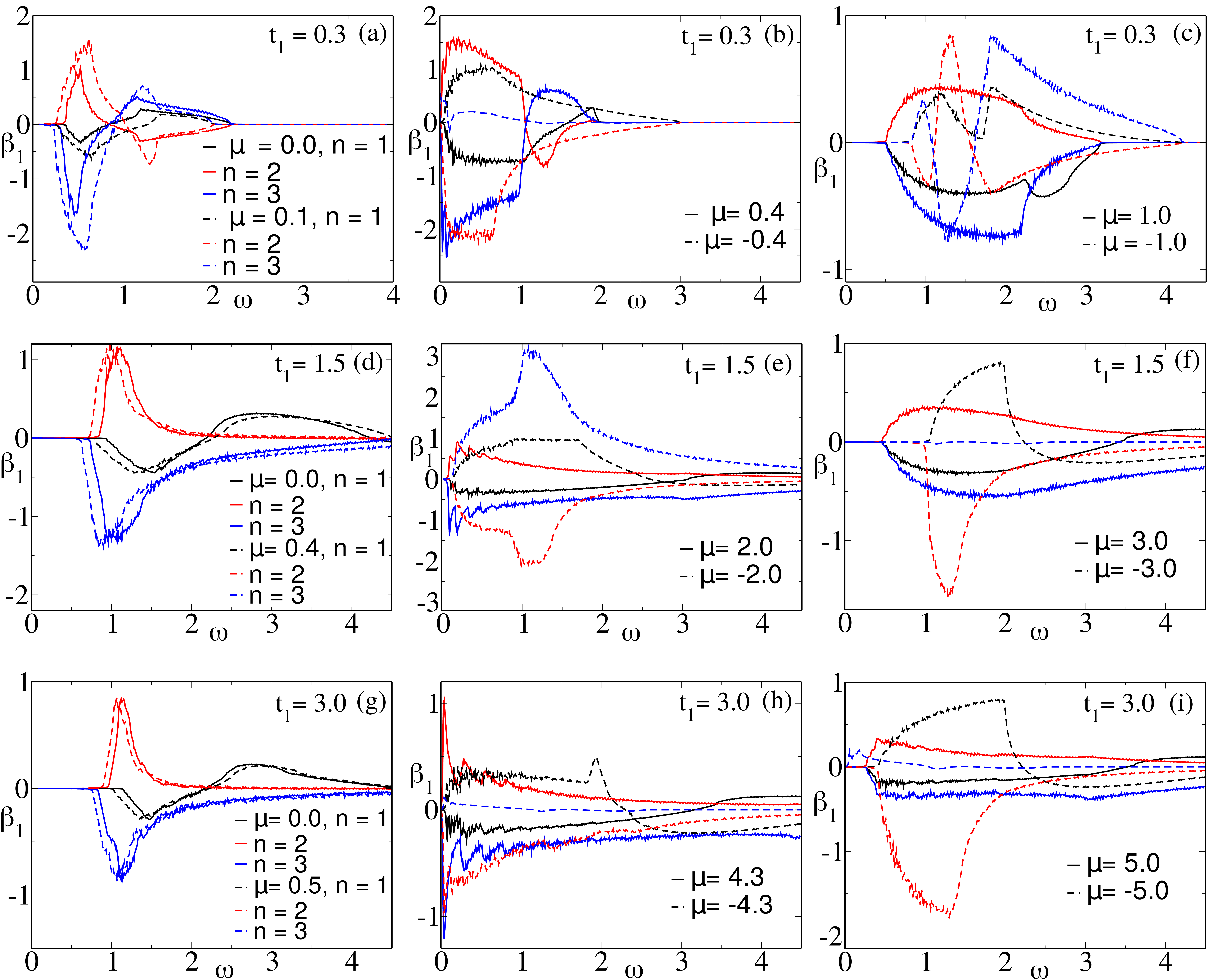} 
\caption{We show the CPGE trace  (Eq.~(\ref{beta})), denoted by $\beta_1$, for type-I phases in (a), (b), (c) with $(t_1,t_2)=(0.3,0.25)$, hybrid phases in (d), (e), (f) with $(t_1,t_2)=(1.5,0.25)$,  and type-II phases in (g), (h), (i) with $(t_1,t_2)=(3.0,0.25)$, considering model I (Eqs.~(\ref{single_WSM}), (\ref{double_WSM}), and (\ref{triple_WSM})). The non-degenerate WNs appear at $E_{\pm}=\pm \sqrt{2} t_1$; $|\mu| < |E_{\pm}|$ for (a), (d), and (g); $|\mu|\approx |E_{\pm}|$ for (b), (e), and (h); $|\mu|>|E_{\pm}|$ for (c), (f), and (i). 
The quantization is most clearly visible in (b) for type-I single WSM when $\mu$ is kept close to the WN's energy $E_+=\sqrt{2} t_1$. With increasing the tilt, the quantization  is     
gradually lost and eventually disappears for type-II WSMs.  For $\mu$ away from the WN's energy, the quantization is completely lost. Most importantly,  $\beta_1$ exhibits quantization to integer values for the hybrid phase in accordance with the chirality of the activated WNs. 
The CPGE trace diminishes significantly for triple WSM while for double WSM, it becomes larger than that for the single  
WSM. The parameters used here are as follows: $t_2=0.25$, $(\phi_1,\phi_2)=(\pi/4,\pi/2)$. The CPGE trace $\beta_1$  is measured in the  unit of $e^3/h^2$. We choose a $k$-mesh of $(300)^3$ points
for our numerical calculations to minimize  finite size effects. 
}
\label{fig:CPG_model1} 
\end{figure*}

We now calculate the trace of  the CPGE   
in the linearized, untilted, isotropic WSM, described by the ${\bm k}\cdot {\bm \sigma}$ model for a single WN. One can show that the 
CPGE trace measures the Berry flux penetrating through a closed surface \cite{de2017quantized}, resulting in a quantized CPGE response proportional to the Chern number ${\mathcal C}$ of the activated 
WN. The frequency windows within which the quantization is observed are  dependent on chemical potential $\mu$. The $\delta$-function 
accounts for the  optical selection rule  that essentially determines the quantization window $2|E'_-|<\omega < 2 |E'_+|$ with $E'_{\pm}=E_{\pm}-\mu$ where  $E_{\pm}$ is the  WN's energy at $k_z =\pm \pi/2$. The quantization is typically lost for 
$\omega > 2 |E'_+|$ where two WNs contributes with opposite sign in the Berry flux.

\begin{figure*} [ht] 
\includegraphics[width = \textwidth]
{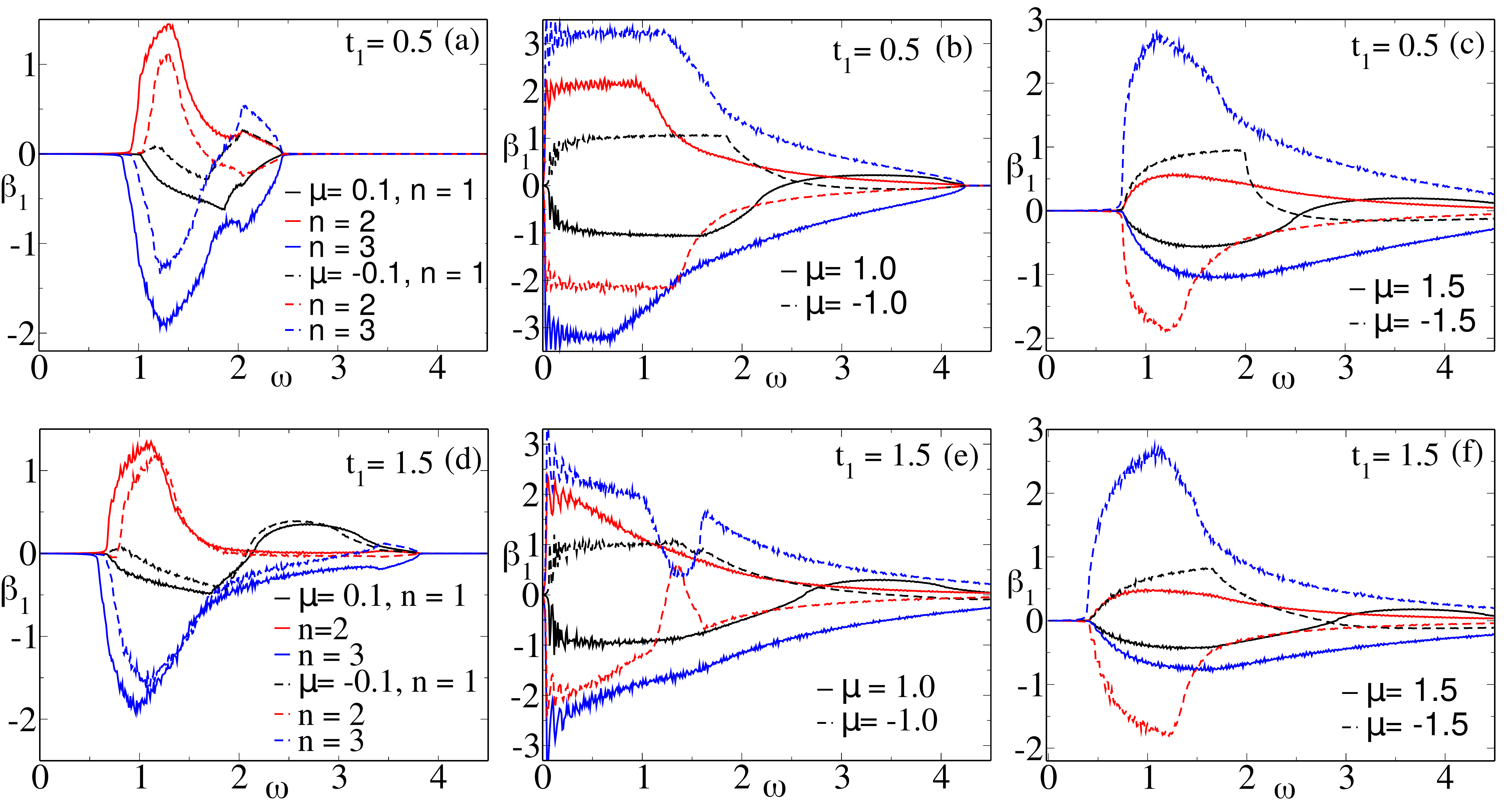}
\caption{We show the CPGE trace $\beta_1$ (Eq.~(\ref{beta})) for model II (Eqs. ~(\ref{single_WSM}), (\ref{double_WSM}), and  (\ref{triple_WSM}) combined with Eq.~(\ref{modelII_n0})) in type-I phase with $t_1=0.5$ (a), (b), (c) and in type-II phase with $t_1=1.5$ (d), (e), (f). The non-degenerate WNs appear at $E_{\pm}=\pm  t_2$; $|\mu| < |E_{\pm}|$ for (a), and (d); $|\mu|\approx |E_{\pm}|$ for (b), and (e); $|\mu|>|E_{\pm}|$ for (c), and (f).
The quantization is clearly visible for type-I phase (b) when $\mu$ is kept close to the WN's energy $E_{\pm}=\pm t_2$. On the other hand, the qunatization  is gradually lost when $\mu$ is away from the WN's energies in the type-II phase.   Interestingly, 
the frequency window, within which CPGE trace becomes quantized, diminishes with increasing the topological charge of the   
 WSM. The parameters used here are the following: $t_2=0.25$, $(\phi_1,\phi_2)=(\pi/4,\pi/2)$. We choose a $k$-mesh of $(300)^3$ points
for our numerical calculations as adopted in Fig.~\ref{fig:CPG_model1}. }
\label{fig:comparison_CPG_typeII}
\end{figure*}

We now discuss the symmetry requirements in order to observe a CPGE response.  The 
tensor $\beta_{ij}$ acquires finite value  if the inversion symmetry is broken.  On the other hand, in complete absence of all the mirror symmetries, the system possesses finite diagonal components of $\beta_{ij}$. In the chiral WSMs, with WNs appearing at different energies,  where inversion and all mirror symmetries are broken, the trace of the CPGE ${\rm Tr}[ \beta(\omega)]$ can show a quantized response. To be more precise, the quantization of CPGE at two opposite plateaus is directly related to the Chern number of the activated WN as noted for TRS broken WSM in Ref.~\onlinecite{de2017quantized}. This picture is modified in an non-trivial way for TRS invariant WSM where CPGE is not quantized at two opposite plateaus even though activated WNs have opposite Chern numbers or chiralities \cite{sadhukhan21,sadhukhan21b}.  
For TRS broken mWSMs, the CPGE trace is expected  to be quantized at higher magnitudes, as compared to single WSMs, in accordance with the higher topological charge associated with the activated WNs. The above predictionw are based on low-energy mWSM Hamiltonians. 
We  below extensively analyze the lattice model I and model II where we denote ${\rm Tr}[ \beta(\omega)]$ as $\beta_1$ to check the validity of this prediction.


We first analyze the results on CPGE trace for model I supporting the additional hybrid mWSM phase. We depict the CPGE trace $\beta_1$ for the type-I phase in Figs.~\ref{fig:CPG_model1} (a), (b), (c), hybrid phase in Figs.~\ref{fig:CPG_model1} (d), (e), (f), and type-II phase in Figs.~\ref{fig:CPG_model1} (g), (h), (i). We consider $|\mu|<|E_{\pm}|$  for Figs.~\ref{fig:CPG_model1} (a), (d), and (g); $\mu\approx E_{\pm}$  is chosen  in    Figs.~\ref{fig:CPG_model1} (b), (e), and (h); $|\mu|>|E_{\pm}|$ are shown in Figs.~\ref{fig:CPG_model1} (c), (f), and (i). In this way, we explore the behavior of $\beta_1$ in different phases as well as the changes in the response  with respect to changing $\mu$.

It is evident from Fig.~\ref{fig:CPG_model1} (b) that  quantization is visible for type-I single and double WSM  where $\mu=-0.4$ is kept close to the WN energy $E_-=- 0.42$.  Similar results are also observed for hybrid mWSM when $\mu$ is kept close to type-I WN at energy $E_-$.
Interestingly, for $\mu $ being  close to $E_+$, the CPGE trace does not show  quantization irrespective of the type and topological charge of the WSMs.  
The double (triple) WSM indeed 
shows twice (thrice) the CPGE trace as compared to that for the single WSM in the hybrid phase only within a certain frequency window (see Fig.~\ref{fig:CPG_model1} (e)). 
However, this frequency window  reduces from single WSM to triple WSM. Interestingly, for triple WSMs, the CPGE trace is not always found to be more pronounced than the single and double WSMs' one while double WSMs show a much more pronounced response than the single WSM one in most of the instances (see Figs.~\ref{fig:CPG_model1} (a), (b), (d), (e), (g) and (h)).  For $\mu \approx 0$ i.e, chemical potential is set around the midway between two WN's energies $\mu\approx(E_+ +E_-)/2$, the CPGE trace  exhibits peak like structure as most prominently  visible for double and triple WSM (see Figs.~\ref{fig:CPG_model1} (a), (d), and (g)). On the other hand, when $\mu$ is set outside the WN's energies, the peaks become flattened yielding non-zero response in a larger $\omega$-range (see Figs.~\ref{fig:CPG_model1} (c), (f), and (i)). For type-I WSMs, the CPGE trace
vanishes for higher frequency (see Fig.~\ref{fig:CPG_model1} (b)).
In contrast, $\beta_1$ remains  non-zero at higher frequency for type-II WSMs (see Fig.~\ref{fig:CPG_model1} (h)). The CPGE trace for the hybrid phase  exhibits a mixed behavior where quantization is observed as discussed above (see Fig.~\ref{fig:CPG_model1} (e)). The frequency window to observe finite CPGE trace depends on $|E'_{+}-E'_{-}|$.


Next, we discuss the signatures of the CPGE trace at certain frequency values. In Fig.~\ref{fig:CPG_model1} (a), one can observe that the CPGE trace
acquires non-zero value above the threshold frequency $\omega =\omega_t> 0.36$ for 
the single WSM when $\mu=0$. The CPGE traces show maximum magnitude roughly at $\omega=\omega_m \approx 0.55$. 
These frequencies can 
be understood from the selection rules $\delta(\hbar\omega - E_{\bm k,12})$, combined  with $\mu=E_{\bm k,1}$ valence and $\mu=E_{\bm k,2}$ conduction band energies, that   determine the optically activated momentum surface. To understand it more easily,
we consider $k_x=k_y=0$ to compute the valence and conduction band energies $E_{k_z,1}$ and 
$E_{k_z,2}$, respectively,
corresponding to the momentum mode $k_{z}$
such that $\mu=E_{k_z}$.
The minimum frequency satisfying the   selection rule $\omega=|E_{k_z,1}- E_{k_z,2}|=|E_{k_z,2}- \mu|$,  around the WN of positive energy $E_+$, is responsible for $\beta_1$ acquireing non-zero values at small $\omega$.  On the other hand,  the selection rules $\omega=|E_{k_z,1}- E_{k_z,2}|=|\mu- E_{k_z,1}|$,  in the vicinity of WN of negative energy $E_-$, qualitatively estimates the frequency at which  $\beta_1$ exhibits a peak. One can also determine the cut-off frequency $\omega_c$ from ${\rm max}\{|\mu- E_{k_z,1}|,|\mu- E_{k_z,2}|\}$ above which CPGE trace vanishes. 
For double and triple WSMs, the non-linearity in the dispersion   is combined with the selection rule to yield
a different set of frequencies $\omega_{m,t}$ as compared to single WSM for $\mu$ being in the proximity to the symmetric position $(E_{+}+E_{-})/2$.

Using the above selection rule for lowest frequency
$\omega =|E_{k_z,1}- E_{k_z,2}|$, one can successfully anticipate the non-zero CPGE response
after a certain threshold frequency in all the cases shown in Fig.~\ref{fig:CPG_model1}.
This is furthermore evident from the fact that CPGE trace attains non-zero value for any finite frequency when $\mu$ is set close to any one of the  WN's energy. 
The CPGE profiles, for $\mu$ being above and below  the WN's energies, are quite different from each other and can be partially explained by the frequency selection rules. The peak structure vanishes there for type-II phases as the selection rules are substantially modified by the tilt as compared to the type-I and hybrid phases. In the hybrid phase, the quantization window is largest (smallest) for single (triple) WSM. As discussed above the optically activated momentum surface, around the activated WN, shrinks with increasing the topological charge.  The 
additional non-linearity in the dispersion 
for mWSMs results in such deviation from quantization to show up more quickly than single WSMs.

We now demonstrate the CPGE trace for the model II that only supports either type-I or type-II phase, as shown in Fig.~\ref{fig:comparison_CPG_typeII}. Here, we adopt the same presentation scheme as followed in Fig.~\ref{fig:CPG_model1} except missing the hybrid phase that no longer exists for model II.  
The frequency window for the quantization increases to non-zero value from zero when $\mu$ is increasing from the symmetric $\mu=0$ to maximally asymmetric $\mu=E_{\pm}$. This behavior is observed for all the mWSMs in the type-I phase (see Fig.~\ref{fig:comparison_CPG_typeII} (a) and (b)). The CPGE trace deviates from its quantized behavior for $|\mu| >|E_{\pm}|$ as shown in Fig.~\ref{fig:comparison_CPG_typeII} (c); however, single WSMs continue to show quasi-quantized response close to $+1$. In the type-II phase, the CPGE trace is found to exhibit quasi-quantization for single WSM when 
$\mu \approx E_{\pm}$ (see Fig.~\ref{fig:comparison_CPG_typeII} (e)).
The frequency window for quantization is confined between  $2E'_-<\omega <2E'_+$ with $E'_{\pm}=|E_{\pm}-\mu|$.
We do not find quantized response in any of the WSMs when $\mu$ is away from $E_{\pm}$ (see Fig.~\ref{fig:comparison_CPG_typeII} (d) and (f)).  For the type-II phase, the CPGE shows finite response in a larger frequency domain as compared to the 
type-I phase.

We now analyze the frequency window within which the CPGE trace acquires finite and quantized values for model II. From the frequency selection rule  $\delta(\hbar\omega - E_{\bm k,12})$ combined  with $\mu=E_{\bm k,1}$ or $\mu=E_{\bm k,2}$ as demonstrated before, one can easily obtain the 
threshold frequency $\omega_t$ and
cut-off frequency $\omega_c$ between which the  CPGE trace becomes quantized 
(see Fig.~\ref{fig:comparison_CPG_typeII} (b)). The quantization for single (double and triple) WSM  can be qualitatively described by the above analysis. In general, the mWSMs 
deviate from quantization earlier than the single WSM case as the optically activated momentum surface shrinks due to additional non-linear terms. 
For $\mu$ outside the energy window between two WNs, the quantization is lost for mWSMs. However, the minimum frequency above which the CPGE trace acquires non-zero value can be 
understood from the selection rule $\omega=\mu-E_{\pm} \approx 0$ for $\mu \approx E_{\pm}$.
For the type-II  case, the estimation of  quantization window, as observed for single WSM, becomes even more complex.

Furhtermore, we compare the results for the CPGE trace between model I and II.  Even though in both the models WN's energies $E_+$ and $E_-$ are substantially separated from each other, the quantization is very prominently visible only for model II. The TRS is broken in both of the models while the specific details of the tilt term, breaking the PH symmetry, can influence the CPGE response to a great extent. Before we present the dissimilarities in CPGE for both the models, the common observations are the following:  The CPGE trace, estimated around a given chemical potential $\mu$, reverses its sign between single/triple and double WSMs. This can be explained from  the lattice model  as the Chern number of the activated WN, when $\mu$ is set close to the corresponding WN energies,
dictates the quantized profile of the CPGE trace.  The other noticeable similarity is that in type-II phases for both models, the CPGE traces slowly vanish with frequency. 
The marked differences between the behavior of the CPGE trace in model I and II are the following:
The  hybrid phase can only show quantization for all three WSMs in the case of model I while type-I (type-II) phase show  quantization (quasi-quantization) in the case of model II. The anti-symmetric nature (i.e., $\beta_1$ just reverses its sign under $\mu\to -\mu$) of the quantized CPGE trace, as clearly observed for model II in type-I phase,  is completely washed  out for model I when $\mu$ is set close to two WNs of opposite chiralities. Surprisingly, the triple WSM is found to show weak response in most of the instances for model I unlike to the model II where it exhibits the most pronounced response as compared to single and double WSMs. 
Overall, model II shows  a quantization of the response more clearly  for single, double and triple WSMs in general.


Additionally, we try to plausiblize the numerical results from generic arguments. We first refer to  the CPGE  formula given in Eq.~(\ref{beta}) where the summands in the  ${\bm k}$-sum can be decomposed into two parts namely,
$\Delta f_{\bm k,12}  \Delta v_{{\bm k},12}^i\Omega_{{\bm k},i}$, being independent of $\omega$ but dependent on the PH term i.e, tilt term  and the remaining part 
$ \delta(\hbar\omega - E_{\bm k,12})$ only depending on $\omega$. In particular,  $\Delta f_{\bm k,12}$ is the only part including the tilt term. The bare energy, without the tilt term, becomes crucial in 
determining the frequency characteristics of the CPGE trace via $ \delta(\hbar\omega - E_{\bm k,12})$. Once the optically activated momentum surface comes into play for a certain range of frequency, the tilt term imprints its effect through the factor $\Delta f_{\bm k,12}$.  The anisotropic non-linear dispersion of mWSM 
significantly affects the magnitude of quantization through the factor $\Delta v_{{\bm k},12}^i\Omega_{{\bm k},i}$.  
Therefore, for model I and model II, the factor $\Delta v_{{\bm k},12}^i\Omega_{{\bm k},i}\delta(\hbar\omega - E_{\bm k,12})$ yields identical contributions while the $N_0$ term significantly changes the profile  as manifested by $\Delta f_{\bm k,12}$. This clearly suggests the importance of the $N_0$ term in determining the CPGE trace.

Importantly, the quantization windows for mWSMs are less as compared to the single WSMs for both the models.
We note another interesting point in Fig.~\ref{fig:CPG_model1} (h) and \ref{fig:comparison_CPG_typeII} (e)
that CPGE shows oscillatory behavior irrespective of its quantization. The oscillatory nature becomes more evident for type-II mWSMs as compared to the single WSM referring to the fact that non-linear band bending in type-II phase play crucial roles. Such oscillations are also noticed for quantized CPGE  in  multi-fold fermions \cite{flicker18}. We additionally note that the noise in the CPGE response neither qualitatively modify the profile of CPGE nor quantitatively alter its magnitude.     

 
\begin{figure*} [ht] 
\centering
\includegraphics[width=\textwidth]
{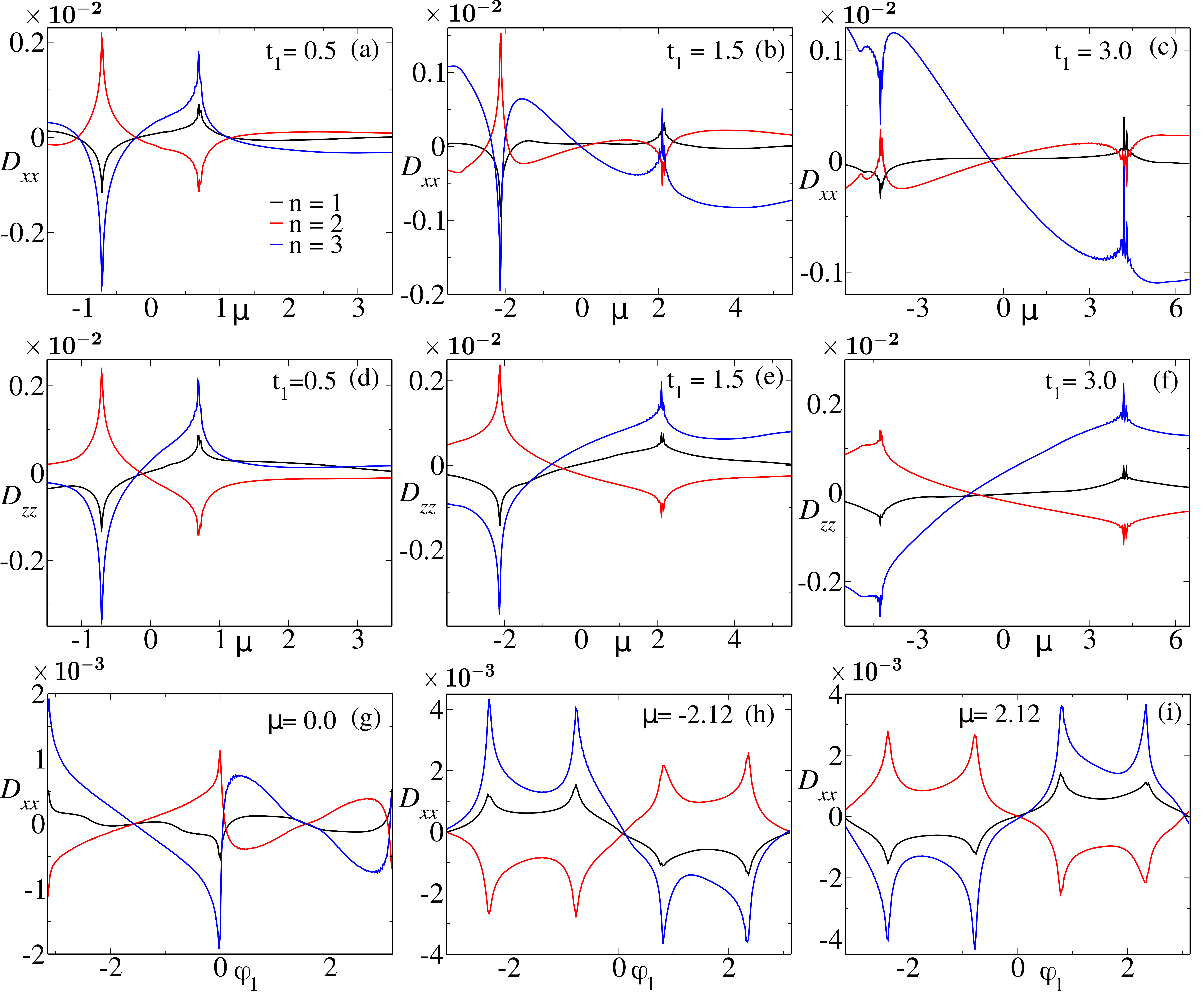}
\caption{The BCD response $D_{xx}$  and $D_{zz}$, following Eq.~(\ref{bcd_density}) for model I, are shown in (a), (b), (c) and (d), (e), (f), respectively. (a), and (d): type-I phase with $t_1=0.5$; (b), and (e): hybrid phase with $t_1=1.5$; (c), and (f):
type-II phase with $t_1=3.0$. 
The common feature, observed for $D_{xx}$ and $D_{zz}$, is that both the WNs contribute almost equally for the type-II phase as compared to the type-I and hybrid phases where the
WN at negative energy $E_-$ contributes significantly. The lower panel (g), (h) and (i) depict the BCD response $D_{xx}$ in the hybrid phase as a function of $\phi_1$ for $\mu=0$, $\mu=E_-=-2.12$ and $\mu=E_+=2.12$, respectively. For $\mu=\pm 2.12$, the BCD response acquires 
substantial contributions around $\phi_1=\pm \pi/4$, $\pm 3 \pi/4$. 
For all the above cases, the responses become more pronounced for higher values of topological charge.  
The parameters used in (a)-(f) are the following: $t_2=0.25$, $(\phi_1,\phi_2)=(\pi/4,\pi/2)$. Al other parameters are the same  as above except $t_1=1.5$  in (g)-(i). Notice that the BCD is measured in the units of lattice constant. }
\label{fig:bcd_modelI}
\end{figure*}
 

\begin{figure} [ht] 
\centering
\includegraphics[width = \columnwidth]
{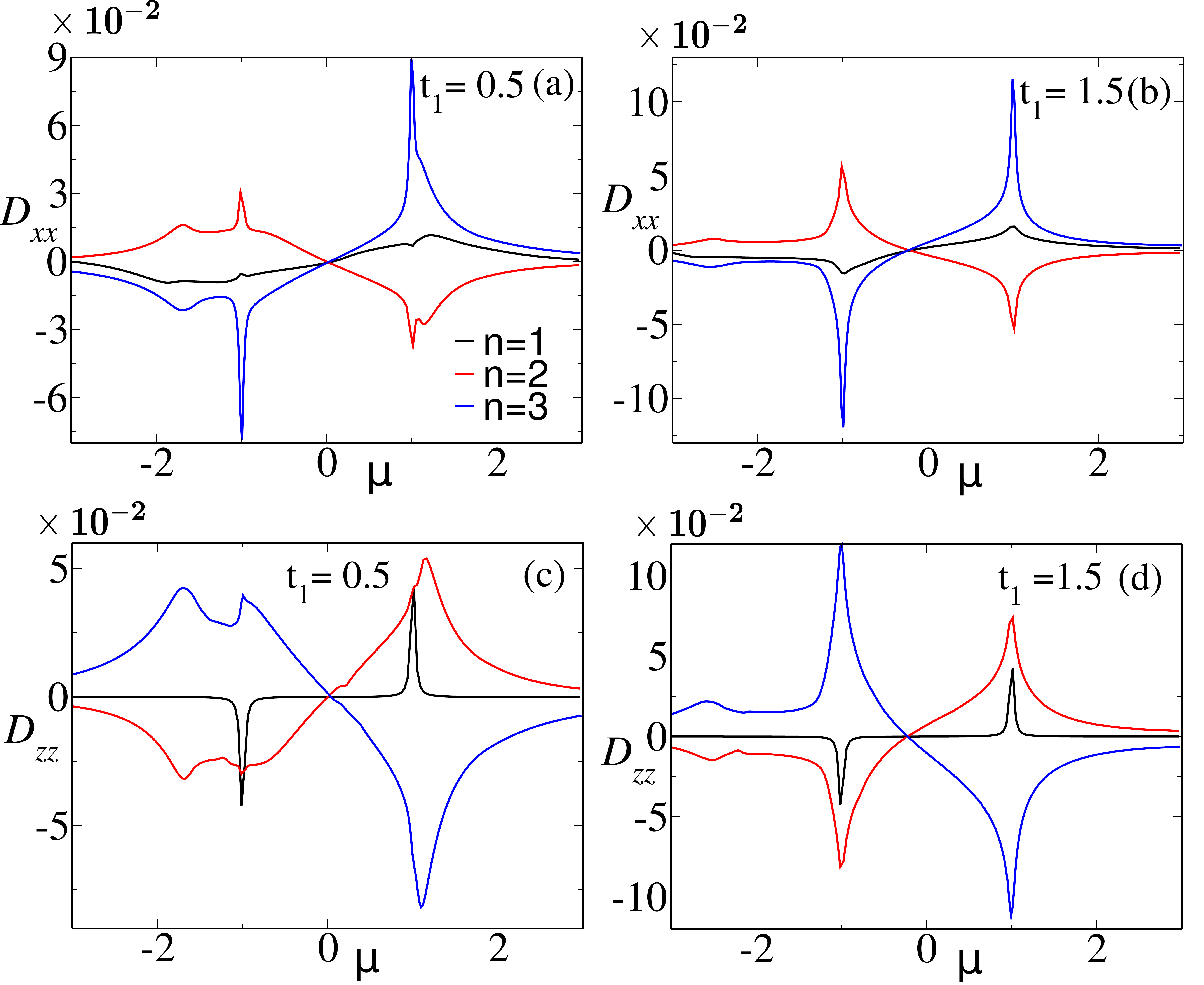}
\caption{We here depict the BCD responses $D_{xx}$   and $D_{zz}$, following Eq.~(\ref{bcd_density}) for model II,  in (a), (b) and (c) and (d), respectively.
(a), and (c): type-I phase with $t_1=0.5$; (b), and (d): type-II phase with $t_1=1.5$. 
Noticeably, $D_{zz}$ for single WSM becomes vanishingly small except at the WN energies $\mu=E_{\pm}=\pm t_2$. This is in contrast to $D_{xx}$
which become significantly diminished only around $\mu=0$.  
The type-I (type-II) double and triple WSMs  with  show (do not show) double peak or dip structure for $\mu<0$.  }
\label{fig:bcd_modelII}
\end{figure}



\section{second order response: Berry curvature dipole (BCD)}
\label{bcd}

Having discussed the non-trivial effects of the PH symmetry breaking tilt term in the Fermi distribution function, we next analyze its consequences on the  Fermi surface properties by investigating the BCD response.  
The DC photocurrent  in  inversion asymmetric systems can lead   to a transverse anomalous velocity $e {\bm E}\times {\bm \Omega_k}$, associated with the Berry phase, when an external electric field ${\bm E}={\rm Re}(E e^{i\omega t})$ is present. This in principle results in an anomalous quantum  Hall effect without any external magnetic field. Interestingly, the first moment of the BC, namely, the BCD is found to be responsible to give rise to the non-linear Hall conductivity $\chi_{abc}$ such that $J_a= \chi_{abc} E_b E^*_c$. We note that this is a diffusive transport phenomena. 
Following the  relaxation time approximation in the Boltzmann equation,
$\chi_{abc}$ is found to be \cite{sodemann15,Zeng21}
\begin{equation}
 \chi_{abc} = \epsilon_{adc} \frac{e^3\tau}{2(1+i\omega \tau)}\int (\partial_b f_{\bm k}) \Omega_{{\bm k},d} d{\bm k}
\label{eq:BCD1}
\end{equation}
where $\epsilon_{adc}$ denotes Levi-Civita symbol and $\Omega_{{\bm k},d}$ represents the $d$-th component of BC.  In the momentum space the BCD thus takes the form 
\begin{equation}
D_{bd}=\sum_m \int d\bm{k} \frac{\partial f_{\bm{k}}}{\partial k_b}\Omega^m_{\bm{k},d}. 
\label{eq:BCD2}
\end{equation}
Here $\sum_m$ refers to the summation over filled bands. It is noteworthy that BCD  is even  under time reversal while BC is odd. Therefore, one anticipates a finite BCD response even for time reversal symmetric systems. Equation (\ref{eq:BCD2}), without loss of generality can be rewritten in the  following form, 
\begin{equation}
    D_{bd}=\int d \bm{k} v_{{\bm k},b}  \Omega_{\bm{k},d} \frac{\partial f_{\bm{k}}}{\partial \epsilon_{\bm{k}}} = \int d \bm{k}\mathcal{D}_{bd} \frac{\partial f_{\bm{k}}}{\partial \epsilon_{\bm{k}}}
    \label{bcd_density}
\end{equation} 
where $\mathcal{D}_{bd}= v_{{\bm k},b}  \Omega_{\bm{k},d}$ refers to  the BCD density \cite{Zeng21}. We are interested in the zero 
temperature limit where  ${\partial f_{\bm{k}}}/{\partial \epsilon_{\bm{k}}}$ is replaced by the Fermi surface configuration $-\delta(\epsilon_{\bm{k}}-\mu)$.

We would now like to comment on the symmetry constraint to have a non-zero BCD response. The off-diagonal BCD response $D_{bd}$ with $b\neq d$ is only found to be non-zero 
if the system preserves two mirror symmetries $M_b$ and $M_d$
simultaneously such that $D_{bd}$ becomes an even function of ${\bm k}=(k_i,k_j,k_l)$ with $M_d{\mathcal H}(k_b,k_d,k_l)(M_d)^{-1}={\mathcal H}(k_b,-k_d,k_l)$  \cite{zhang18,Yu19,Zeng21}. On the other hand, the diagonal element are expected to contribute  without the above symmetry requirement. Considering the single node low-energy model of WSMs, it has been found that  the off-diagonal component $D_{bd}$ vanishes while the diagonal components $D_{bb}$ continue to exist in presence of a finite tilt. However, the BCD response in the 
mirror symmetry protected inversion broken lattice models of WSMs do not agree with the predictions from low-energy models \cite{Zeng21}. We therefore, investigate the lattice model I and II to predict the possible   experimental observations associated with TRS broken WSMs.


 We first investigate $D_{xx}$ and $D_{zz}$  for the type-I phase of model I in Figs.~\ref{fig:bcd_modelI} (a), and (d), the hybrid phase in 
Figs.~\ref{fig:bcd_modelI} (b), and (e) and the 
type-II phase in  Figs.~\ref{fig:bcd_modelI} (c), and (f), respectively. The dependence of $D_{xx}$ inside the hybrid phase as a function of $\phi_1$  are shown in Figs.~\ref{fig:bcd_modelI} (g), (h) and (i) for $\mu=0$, $-2.12$ and $2.12$, respectively. We next probe $D_{xx}$ and $D_{zz}$  for the type-I phase of model II in Figs.~\ref{fig:bcd_modelII} (a), and (c) and 
the type-II phase in  Figs.~\ref{fig:bcd_modelII} (b), and (d), respectively.
As we discussed above the lattice Hamiltonian for WSMs (Eqs.~(\ref{single_WSM}), (\ref{double_WSM}) and (\ref{triple_WSM})) break mirror symmetries such as $M_i^{\dagger} H(k_x,k_y,k_z)M_i \neq H(\alpha k_x, \beta k_y, \delta k_z)$ with $(\alpha,\beta,\delta)=(-1,1,1)$, $(1,-1,1)$ and $(1,1,-1)$ for $i=x$, $i=y$, and $i=z$, respectively. Therefore, we do not see any cross term $D_{bd}$ contributing to the non-linear transport rather only diagonal 
terms $D_{bb}$ become non-zero. The most significant contribution is expected to come from the vicinity of
the WN i.e., when $\mu$ is close to WN energies. With increasing topological charge, the BC enhances that in turn causes the BCD to grow. 
Another interesting feature is  that $D_{zz}$ and $D_{xx}$ behave differently in general, and
this is most prominently visible for double and triple WSMs.  This is due to the fact that $v_{{\bm k},x}$ and  $\Omega_{{\bm k},x}$ are significantly different from $v_{{\bm k},z}$, $\Omega_{{\bm k},z}$. 
On the other hand, $D_{xx}$
and $D_{yy}$ behave identically due to the similar structure of velocity and BC.

Now examining  Fig.~\ref{fig:bcd_modelI} (a)-(f), we can clearly observe that the
BCD response is most pronounced at $\mu \approx E_{-}=-\sqrt{2}t_1$ while the BCD contribution is smaller for $\mu$ close to the WN of positive energy. This is in spite of the fact that the magnitude of BC remains the same around both the WNs.
We note that the WN at energy $E_{-}$ is type-I for the hybrid phase. For $\mu$ away from the WN's energies, the BCD response diminishes due to the fact that the BC and BCD both become significantly reduced.
Interestingly, the sign of the topological charge is  reflected in the diagonal BCD response.
For $\phi_1=\pm \pi$ and $0$, one can find a pronounced peak and dip as the Fermi surfaces, associated with the two degenerate WNs at $\mu=0$, interfere constructively provided the fact that $v_{{\bm k},x} \to -v_{-{\bm k},x}$ and ${\bm \Omega}_{{\bm k},x} \to -{\bm \Omega}_{-{\bm k},x}$
(see Figs.~\ref{fig:bcd_modelI} (g)-(i)). On the other hand, for $\phi_1=\pm \pi/2$, the WNs are separated from each other in the energy space. As a result, their  associated Fermi surface contributions becomes vanishingly small at $\mu=0$. The leads to a substantial reduction of BCD response. The sign of the response depends on whether the positive (negative) chiral WN  approaches
$\mu=0$ from below (above) or above (below).  We note that a given WN becomes more tilted when $\phi_1$ approaches to $0$ and $\pm \pi$ from $\pm \pi/2$.  We encounter another instance where the Fermi surface of a single WN leads to a BCD response as depicted for $\mu=2.12$. Here, 
type-I (type-II) WNs appear for $\phi_1=-\pi/4$ and $3\pi/4$ ($\phi_1=-3\pi/4$ and $\pi/4$).
The response drops significantly for $\phi=0$, $\pm \pi$ when the degenerate WNs appear at $\mu=0$ far away from $\mu=\pm 2.12$. Therefore, the contribution from the BC drops substantially leading to a suppressed BCD response.

Having explored the combined effect of tilt and gap in model I, we now turn to 
model II as shown in Fig.~\ref{fig:bcd_modelII} for type-I and type-II phases. We here find a secondary peak (dip) in addition to the primary  peak (dip) in the type-I phase  for $D_{xx}$ and $D_{zz}$. This  secondary peak or dip
almost vanishes for the type-II phase.
This is clearly visible for double and triple WSMs. The position of the primary peak or dip is directly given by the gap term $t_2$ itself while the secondary peak is weakly dependent on the tilt term $t_1$ (see Fig.~\ref{fig:bcd_modelII} (a) and (b)).   The anti-symmetric Fermi surface contributions, associated with the individual WNs, interfere destructively at $\mu=0$ leading to the  minimum in the BCD response. On the other hand, with increasing topological charge, the asymmetric nature of Fermi surface is clearly visible in the BCD response  between $\mu <0$ and $\mu >0$. To this end, we focus on  the BCD response $D_{zz}$ where  it vanishes everywhere except at WN's energies in the case of single WSM (see Fig.~\ref{fig:bcd_modelII} (c),and (d)). The asymmetric response is more clearly noticed here and the primary peak or dip locations
differ significantly between type-I and type-II phases. This might be related to the anisotropic non-linear dispersion for double and triple WSMs. Interestingly, the secondary peaks in the type-I phase indicates that the BCD density  $\mathcal{D}_{bb}= v_{{\bm k},b}  \Omega_{\bm{k},b}$ with $b=x,~z$ acquires finite value even away from WNs \cite{Zeng21}.

Interestingly, $D_{zz}$ qualitatively follows $D_{xx}$ for model I while they become different for model II.  The BCD responses, observed  for the type-I phase in model II, only show secondary peaks or dips at certain $\mu$ other than the WN energies. By contrast, there is no prominently secondary peak or dip structures found for model I. Therefore, the Fermi surface properties are significantly modified by the PH term associated with $\sigma_0$ even though the topological characteristics remain unaltered. In addition, the BCD density also changes from model I to model II causing the overall BCD to behave distinctly.   
To be precise, the underlying nature of the model is reflected in the vanishing off-diagonal components of BCD while the sign of the diagonal components at certain $\mu$ is determined by the chiralities of the WNs.


\section{Magnus Hall conductivity (MHC)}
\label{MHC}

\begin{figure*}  [ht]
\includegraphics[width = \textwidth]
{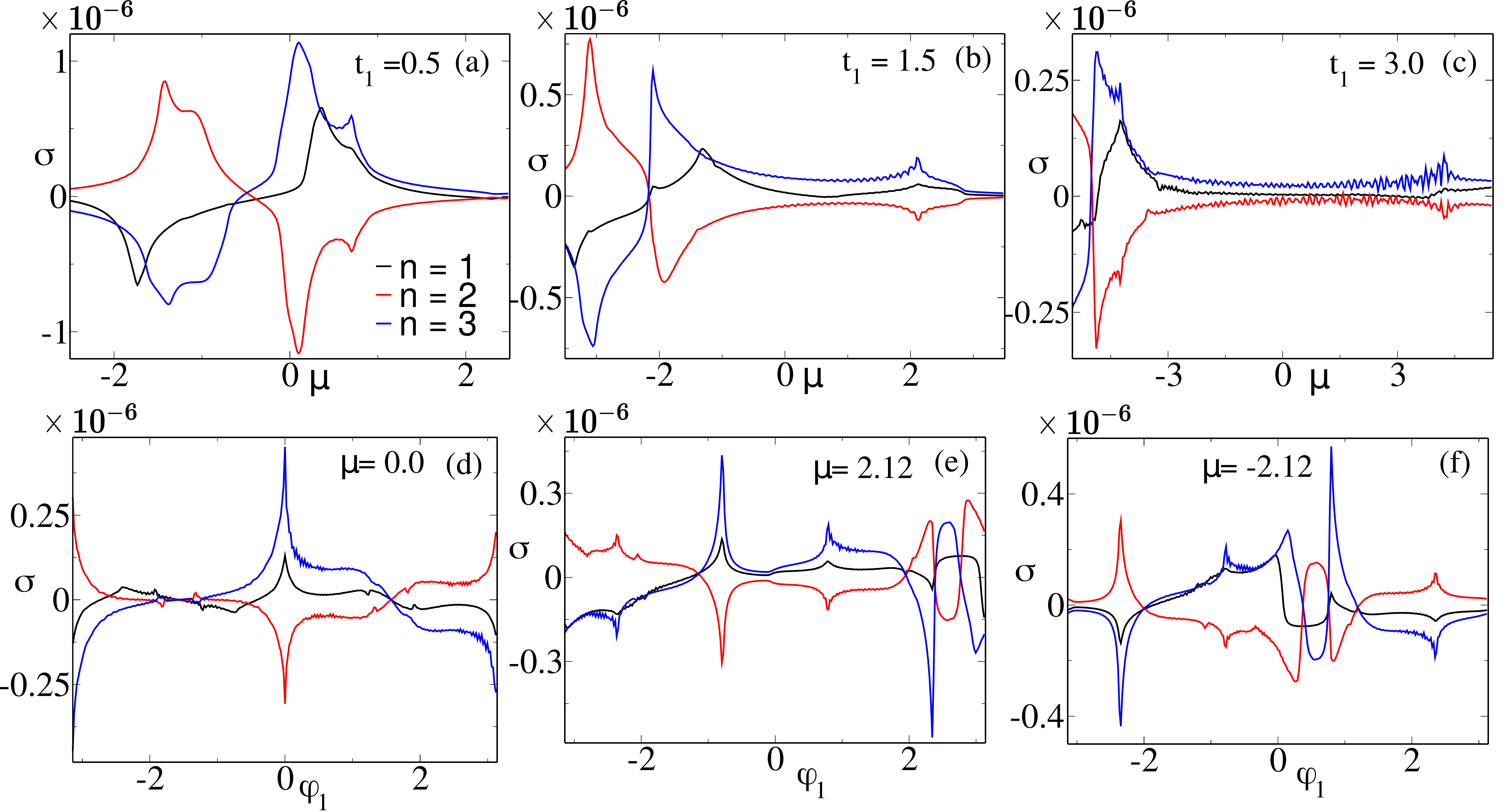}
\caption{The MHC $\sigma$, following Eq.~(\ref{sigma}), in the type-I, hybrid and type-II phases for  model I are depicted 
in (a), (b), (c)  and (d), (e), (f), respectively, 
as a function  of $\mu$ and $\phi_1$. The MHC shows pronounced signature at certain values of  $\mu$ that depend on Fermi surface property of the lattice model. Interestingly, the magnitude of the MHC decreases with increasing the tilt. The MHC exhibits complex behavior with $\phi_1$ as the Fermi surface 
changes substantially there. 
The parameters used for (a)-(c) here are $t_2=0.25$, $t_1=0.5$ (type-I), $t_1=1.5$ (hybrid) and $t_1=3.0$ (type-II), $(\phi_1,\phi_2)=(\pi/4,\pi/2)$. The parameters chosen for (d)-(f) are $t_1=1.5$, $t_2=0.25$, and $\phi_2=\pi/2$. The MHC is measured  in the unit of $e^2/h$.}
\label{fig:mhc_modelI}
\end{figure*}


\begin{figure}  [ht]
\includegraphics[width = \columnwidth]
{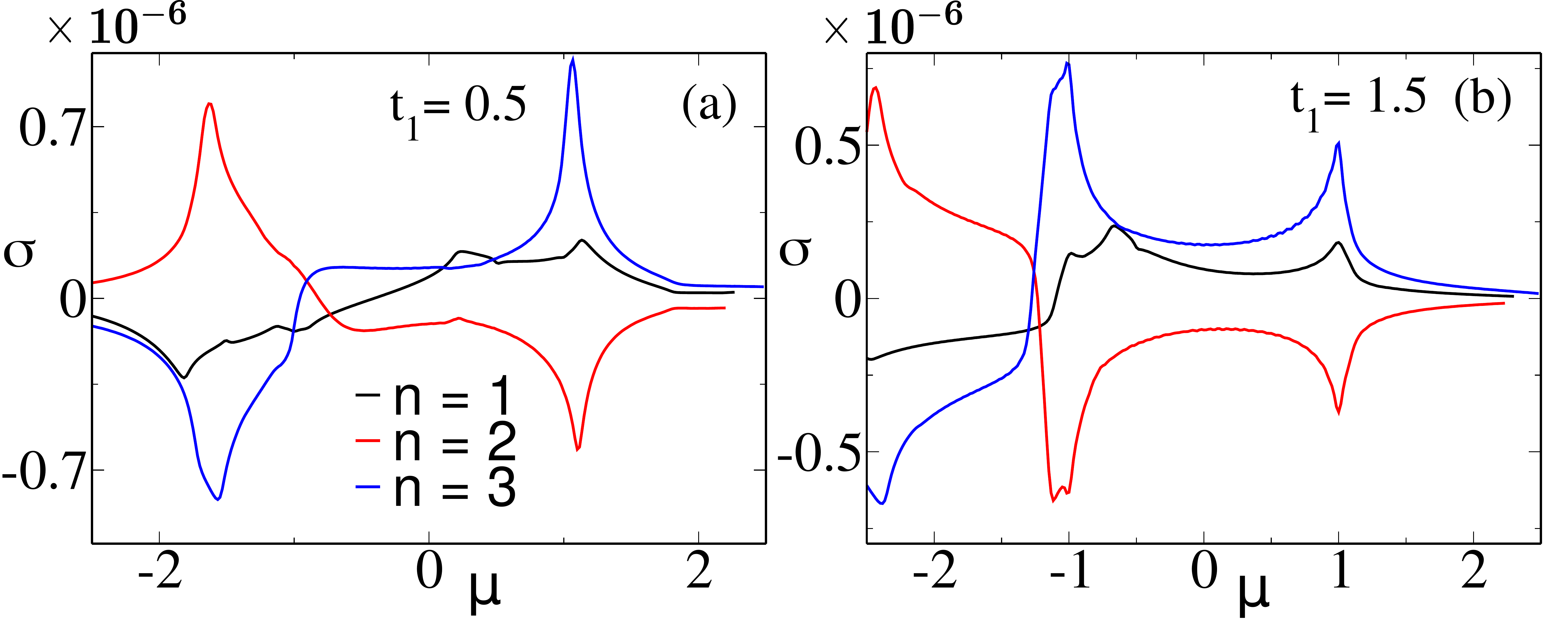}
\caption{The MHC $\sigma$ following Eq.~(\ref{sigma}) for model II, is depicted
with $\mu$ for type-I and type-II phases in (a) and (b), respectively. The magnitude of the MHC remains almost unaltered for the type-II phase as compared to the type-I phase unlike the model I as shown in Fig.~\ref{fig:mhc_modelI}. However, the MHC profile significantly changes with increasing the tilt. The parameters used are the following: $t_2=1$, $t_1=0.5$ for the type-I and  $t_1=1.5$ for the type-II phases.}
\label{fig:mhc_modelII}
\end{figure}


We here briefly discuss the  MHC that is derived in 
the ballistic regimes using Boltzmann transport equation without applying any external magnetic field \cite{papaj19}.  The difference between the gate voltages $U_s$ and $U_d$, associated with the source and drain, $\Delta U= U_s-U_d$ introduces a built-in electric field ${\bm E}_{\rm in}={\bm \nabla}_\mathbf{r} U/e$ ($-e$ is the electronic charge) in the Hall bar with a slowly varying electric potential energy $U(r)$  along the length of the sample. 
With the relaxation time
approximation the steady-state Boltzmann equation is written as \cite{ashcroft1976solid,ziman2001electrons}
\begin{equation}
(\mathbf{\dot{r}}\cdot\mathbf{\nabla_{r}}+{\dot{\bm k}}\cdot{\nabla_{\bm k}})f^{\rm ne}_{{\bm k}}=\frac{f_{\bm k}-f^{\rm ne}_{{\bm k}}}{\tau},
\label{eq_BZf}
\end{equation}
where  the scattering time $\tau$ can be considered independent of momentum ${\bm k}$  and the  equilibrium (non-equilibrium) electron distribution function is denoted by $f_{\bm k}$ ($f^{\rm ne}_{\bm k}$). In the ballistic regime,
the mean free time between two collisions is infinite $\tau \rightarrow \infty$ suggesting the fact that electrons entering from the source traverse (say along $x$-direction) to the drain with positive velocity $v_x >0$ without experiencing a collision  within the length $L$ of the Hall bar. 
Employing the semi-classical equation of motion in the Boltzmann equation without the right hand side collision term and following the ansatz
$ f^{\rm ne}_{\bm k}=f_{\bm k}-\Delta \mu \partial_\epsilon f_{\bm k}$,  one can find  the MHC in response to the external electric field $E_x$ as given by \cite{papaj19,Mandal20,das21}
\begin{equation}
\sigma =- \frac{e^2}{\hbar} \Delta U \int_{v_x>0} d{\bm k}~ \Omega_{{\bm k},z}  \partial_\epsilon f_{\bm k}~,
\label{sigma}
\end{equation}
where $-eL E_x =\Delta \mu$ and ${\bm E}_{\rm in}=E_x \hat{i}$.

It is noteworthy that the MHC  (Eq.~(\ref{sigma})) originates from the Magnus velocity $V_{\rm magnus}=\nabla_{\bm r}{U} \times \bm \Omega $ that can be  thought of a quantum analog of the classical Magnus effect. The Magnus responses can be 
effectively 
considered as a second order coefficient as the built-in electric field $\Delta U$ and external electric field both appear in calculation of currents. As far as the
symmetry requirements are concerned, it has been shown that in  
presence of crystalline symmetries such as, specific $C_2$ and certain mirror symmetries, MHC shows a non-trivial response  
 \cite{Xiao2021}.
More importantly, for the  Magnus responses to become non-zero, the  system must possess finite BC and asymmetric Fermi surface. The MHC in a way 
allows us to scan through the
Fermi surfaces by tuning $\mu$ and investigate the angular distribution of BC within a given Fermi surface. It is worth mentioning that the MHC is found to vanish for low-energy model of untilted mWSMs while  it becomes less pronounced with increasing the topological charge in presence of tilt\cite{das21}.
We shall further examine  these properties below by considering the generic lattice models namely, model I and model II of WSMs.

We analyze the MHC response for type-I, hybrid and type-II phases in 
Figs.~\ref{fig:mhc_modelI} (a), (b) and (c), respectively, for model I. We then show the dependence of the MHC with respect to $\phi_1$ for the hybrid phase with $\mu=0$ and $\pm 2.12$, respectively, in 
Figs.~\ref{fig:mhc_modelI} (d), (e) and (f). 
The behavior of the MHC in model II are demonstrated for type-I and type-II phases  in Figs.~\ref{fig:mhc_modelII} (a) and (b), respectively.
The main contribution is coming from the
$z$-component of BC and Fermi surface properties for the ${\bm k}$-modes having positive quasi-velocity along the transport direction. Therefore, the MHC can 
show higher amplitude at certain $\mu$ 
different from WN energies.
The peak or dip in the MHC moves away from $\mu=0$ as the tilt increases (Fig.~\ref{fig:mhc_modelI} (a), (b), and (c)). This can be naively understood from the fact that the WNs depart from each other with increasing tilt.  The magnitude of MHC decreases with increasing the tilt in contrast to the prediction obtained from low energy mWSM model \cite{das21}.
The signature of WNs, however, appears in the MHC; for example, WNs exist for $\phi_1=0$ at $\mu=0$ resulting in pronounced behavior (see Fig.~\ref{fig:mhc_modelI} (d), (e) and (f)). A similar behavior is also observed for $\mu=\pm2.12$ where WNs appear $\phi_1=\pm \pi/4$ and  $\pm 3\pi/4$.

We now discuss the MHC behavior in the model II where the gap and tilt induce marked effects (see Fig.~\ref{fig:mhc_modelII}). The 
peaks in the
MHC transform into dips while crossing through $\mu=0$ in the type-I phase. This is not observed for the type-II phase as the Fermi surface is modified by the tilt.  The effect of higher topological charge in general can be observed in the larger amplitude of the MHC. Another dissimilarity between MHC results, obtained from model I and model II, is that for $\mu$ being close to a WN at negative energy $\mu\approx E_{-}$,
the amplitude of the MHC enhances significantly  for model I. The MHC responses thus do  not necessarily maximize always around WNs unlike the BCD response. 

\section{discussions}
\label{discussion}

We here compare our findings with the existing results where the non-linear responses are investigated. We start with the CPGE trace for the TRS invariant model \cite{sadhukhan21}. The CPGE trace there is found to be quantized at higher values than the combined topological charge of the activated WNs unlike to the TRS broken case \cite{de2017quantized}. We here focus on the other aspect i.e, PH symmetry instead of TRS symmetry, by which the  property of individual WNs can be tuned. Considering  the general lattice model, embedded with the non-linear and anisotropic dispersion, our study unfolds three effects: Firstly, unique quantization profiles of the CPGE trace in the hybrid phase, secondly distinct signatures when the tilt and gap term are mutually decoupled, and lastly quantization to higher values for mWSMs as compared to single WSMs. Therefore, our work sheds light on the generic properties of the CPGE trace for a TRS broken mWSMs with non-zero chiral chemical potential such that $E_{+}\neq E_-$. The frequency window within which the CPGE acquires quantized values is non-trivially modified by the choice of the PH symmetry breaking terms.

Moving to the next part of BCD induced second order responses, we comment that TRS invariant system, possessing a certain mirror symmetries, show interesting off-diagonal BCD response \cite{Zeng21}. Due to the lack of mirror symmetries, we find only diagonal component to become non-zero. We note that the diagonal components are only found to be non-zero   following the single node low-energy model analysis. However, our findings are similar though the underlying physics is different. It has been shown that BCD
is related to properties on the Fermi surface in the overlapping region between the WNs (around the WNs)  can result in off-diagonal response when the WNs are degenerate (non-degenerate) \cite{Zeng21}. In our model I (II),   we find one (both) of these signatures where gap and tilt terms are coupled (decoupled) form each other.

The substantial BCD response away from the WNs, noticed for model II, might be related to a different  BCD density distribution as compared to the regular BCD density leading to a strong BCD response around the WNs, observed for model I.   
Therefore, even though the off-diagonal terms vanish due mirror symmetry constraints, the essential physics remains the same for the diagonal components. Finally, considering  the low-energy mWSMs, it has been shown that tilt can lead to a finite MHC and the responses become less pronounced with increasing the topological charge   \cite{das21}. 
Interestingly, the findings on the MHC here, obtained from lattice models, are substantially different from results based on the low-energy model similar to the BCD response \cite{Zeng21}.
The MHC is found to be finite even for the untilted case and its magnitude is generically increases with non-linearity in the dispersion. Therefore, one needs to compute the transport coefficients for the lattice model in order to obtain a more realistic picture closer to the experiment.

We would now like to discuss the common symmetry constraints in order to observe these higher order effects.  The transport coefficients such as, CPGE trace, BCD response, and MHC can become non-zero even if the underlying system does not break TRS. However, we consider TRS broken WSMs where the first order effects such as the linear anomalous Hall response can become substantial. We note that the linear anomalous Hall response is not a direct Fermi surface phenomenon, however, it depends on the number of filled bands. As a result, it does not depend on the
derivative of the Fermi function unlike the BCD and MHE. Therefore, from the profile of these responses as a function of  chemical potential, one is able to
distinguish them from the linear anomalous Hall response. On the other hand, the prefactor $\beta_0$ in the CPGE trace can be large causing a quantized signature to be observed as comparison to the much reduced metallic or insulating contributions \cite{sodemann15,Moore10,ni2021giant}. 
It is to be noted that 
in order to observe finite CPGE, the WNs must be non-degenerate. In contrast,  the WNs need not to be non-degenerate to realize finite BCD or MHC.


We would like to comment on the  
experimental realization of the transport coefficients in various 2D and  3D systems. The bulk quantum Hall effect  has already been realized  in quasi-2D systems \cite{Cao2012,Masuda2016,Uchida2017}. 
Interestingly, the recent experiments are not restricted to 2D systems only,  
ZrTe$_5$, HfTe$_5$, and Cd$_3$As$_2$ are examples of 3D systems that 
exhibit a quantum Hall effect \cite{Zhang2018,Liang2018,Tang2019,Galeski2020}.  Apart from the external magnetic field induced quantum Hall effect, the 
non-linear Hall effect, mediated by BCD,  has been experimentally observed in the bilayer non-magnetic quantum material WTe$_2$ \cite{Ma2018},  in a few layers of WTe$_2$ \cite{Kang2019}, and in type-II WSM at room temperature \cite{Kumar2021}.
On the other hand, the  type-I WSM transition monopnictide family such as TaAs, is found to exhibit an interesting CPGE trace 
\cite{ma2017direct,osterhoudt2019colossal};
moreover, CPGE is also extensively investigated for Type-II WSMs where the response is not directly linked to their topological charge \cite{ji2019spatially,ma2019nonlinear}. Very recently, the  chiral multifold semimetals such as, RhSi are found to exhibit 
non-quantized CPGE \cite{ni2020linear}, however, first-principle theoretical studies demonstrate quantization in contrast \cite{Guoqing17,flicker18,Le_2021}.

Given the above developments in the   experiments, it is in principle possible to
use the candidate  double (HgCr$_2$Se$_4$) and triple WSM (Rb(MoTe)$_3$) materials as samples to investigate the non-linear Hall effect. We note that type-I WSMs i.e., TaAs family and type-II WSMs i.e., WTe$_2$ family both break inversion symmetry. We here discuss inversion symmetry broken mWSM where the hybrid phase can be engineered that is yet to be realized in the experiment. However, magnetic doping could be one of the possible approaches through which hybrid phase can be obtained \cite{fyli16}. As far as the experimental set up in 3D is concerned, the 2D systems can be stacked together forming a quasi-2D / 3D structure where multi-terminal Hall measurements can be performed with appropriate gate potentials. However, the exact prediction of material and accurate description of experimental set up is beyond the scope of the present study.

\section{material connections}
\label{material}

Here, we connect our findings, based on TRS broken WSM models, with material studies on WSMs following first-principle  calculations. 
In the double WSM SrSi$_2$, preserving TRS, CPGE response does not show opposite quantization profile for $\mu$ being close to two opposite chiral WNs \cite{sadhukhan21b}. This finding is qualitatively similar to the quantization profile for both of our present models referring to the fact that our study is useful in predicting the CPGE profile in real material. Turning to the BCD,  it has been found that BCD in type-II (MoTe$_2$ family) is more pronounced than type-I (TaAs, NbAs, and NbP family) \cite{zhang18}. In our case, we find similar behavior for model II. The BCD there acquires maximum value away from the WN energies that is also noticed for MoTe$_2$ family. On the other hand,  the MHE is only studied in 2D transition metal dichalcogenides \cite{papaj19,Xiao2021} and 
is yet to be explored using first-principle  calculations in 3D WSM. However, we believe that our study yields a broad picture of second-order responses possible in
theoretical models of mWSMs which are relevant in the context of real material scenarios.

\section{experimental connections}
\label{experiment}

{Considering the recent progress on the experimental side, we here demonstrate a possible route to  realize  the above transport coefficients in practice. Note that 
for TRS broken WSM, the leading order contribution  of order $(\Omega {\rm cm})^{-1}$ would come from the anomalous Hall conductivity \cite{shekhar2018anomalous}. The response discussed in the present work are subleading and can acquire values $O[(\mu \Omega {\rm cm})^{-1}]$-$O[(m \Omega {\rm cm})^{-1}]$. 
It has been observed in multi-fold fermion CoSi that the output voltage signal reverses its sign under the reversal of the input polarization, controlled by quarter wave plates, yielding a direct experimental signature of CPGE
\cite{ni2021giant}. The quantized CPGE signal is not yet experimentally realized to the best of our knowledge even though the first-principle ab initio studies identifies  a clear quantized signal \cite{sadhukhan21b,Guoqing17,flicker18,Le_2021}. We believe that our theoretical findings on CPGE  can be verified in the optical conductivity measurements  using the terahertz  emission spectroscopy \cite{ni2020linear}.
From the  in-phase and out-of-phase photocurrent, the CPGE signal can also be  determined in the multi-terminal device that is thicker than the penetration depth of the light \cite{osterhoudt2019colossal}.
As far as the magnitudes of the CPGE is concerned,  CoSi is shown to exhibit photocurrent of $O(\mu A)$ where the relaxation time is of the order of Femtosecond \cite{ni2021giant}. We comment that if
the relaxation time is larger or comparable with the external pulse width,  the quantization $O(\mu AV^{-2})$ 
is expected to occur in the THz regime \cite{ni2020linear}. Note that the 
anomalous Hall conductivity is insensitive to the polarization of external electromagnetic field from which CPGE can be distinguished even though the second-order response is substantially small. 
}

{On the other hand, the non-linear Hall effect, induced by BCD, is experimentally observed where  the transverse Hall voltage $O(\mu$V) is found to vary quadratically with the longitudinal current $O(\mu$A) for non-magnetic few-layer WTe$_2$ \cite{kang2019nonlinear,ma2019observation}. The BCD, estimated there, are found to be $O(10^{-1}-10^{0})$ nano-meter.  In the present case without TRS and any mirror symmetry, we believe that longitudinal voltage can indicate the existence of the diagonal BCD following   the angle-resolved electrical measurements. We also expect   the longitudinal voltage  to be $O(\mu$V) under the longitudinal current $O(\mu$A) such that the diagonal BCD can become $O(\AA)$.}

{Turning to the MHE, we comment that it has not been experimentally observed yet to the best of our knowledge. However, the 3D generalization of MHC might not be obvious in terms of the Hall-bar experiments that are mostly based on the gating in 2D sample. However, for 3D systems, such gating can be implemented 
in multi-layer structures of WSMs such as, a few
layerd of WTe$_2$ \cite{Kang2019}. In order to obtain the MHC along $y$-axis for 3D system, the motion of the electron is restricted in the two-dimensional $xy$-plane while in the other direction along $z$-axis, the electron's wave-function is localized. This might be obtained by  tuning the gate voltage along the $z$-direction. There could be another way to 
engineer the built-in electric field in the transport plane where the strain  
is introduced only along $x$-direction \cite{bykhovski1993influence,ren2004large,Huang2019}.
In this case, one has to be careful about the fact that the WNs remain unaltered even in the strained case \cite{Heidari20}. The Fermi arc surface states might play an interesting role in MH transport if the electrons move coherently with a well defined velocity under the built-in electric field in the surface of WSM.
The accurate mechanism of 
such movements of electron on the 2D  plane, say top surface in the 3D system, is beyond the scope of the present study. The possible length scale along $x$-direction resulting in finite MHC i.e., transverse voltage along $y$-direction, can be estimated from 
$\lambda_x \sim v_F/ \tau$ where $v_F$ is the characteristics Fermi velocity and $\tau$ is the relaxation time between two successive collisions. The length scale for ballistic transport is found to be $O(nm)$-$O(\mu m)$ as $v_F\sim 10^5-10^6 m/s$ and $\tau\sim O(10^{-15}-10^{-12})s$. Without loss of generality,
the dimensions $L_x \simeq \lambda_x$, and $L_y$ of the system along $x$ and $y$-directions, respectively,
are comparable while $L_z$ can be made larger such that 
$[L_x,L_y,L_z]\approx [O(nm),O(nm),O(\mu m)]$. 
We believe that  MHC
can acquire values $O[(\mu\Omega {\rm cm})^{-1}]$-$O[(m\Omega {\rm cm})^{-1}]$ where the built-in electric potential $\Delta U$ can be  typically of the order of meV. The linear dependence on $\Delta U$ is a clear signature for the  ballistic transport that can be experimentally observed.  
 }

\section{conclusion}
\label{conclusion}

We conclude by assembling our results on the second order transport coefficient, such as CPGE and BCD responses, in a more general way.  Both the above effects are found to be finite for a TRS invariant system where the first order quantum anomalous Hall effect vanishes. Interestingly, a finite CPGE trace requires all the mirror symmetry to be broken in addition to inversion symmetry breaking that  results in non-degenerate WNs. On the other hand, the BCD mediated responses become finite for systems having a certain mirror symmetries. Therefore, the situation becomes complex if the TRS, IS and mirror symmetry are broken which we study in our work. In order to analyze the problem more deeply we consider two types of PH symmetry breaking terms in model I and model II (Eqs.~(\ref{single_WSM}), (\ref{double_WSM}), (\ref{triple_WSM}), (\ref{modelII_n0})). This  allows us to investigate a complex WSM phase namely hybrid phase where one WN is of the type-I and the other 
is of type-II. We consider tight-binding lattice model for mWSM whether the topological charge associated with each WN is larger than unity.  Therefore, our work on one hand, explores the effect of anisotropy and non-linearity of the dispersion and, other hand, sheds light on the influence of tilt and gap factor in the transport properties.

In the case of model I where the effects of the tilt and gap are combined, we find quantized CPGE for the hybrid phase only (see Fig.~\ref{fig:CPG_model1}).  The gap and tilt terms are decoupled in model II, the pronounced quantized CPGE trace within an extended frequency window is found for the type-I phase (see Fig.~\ref{fig:comparison_CPG_typeII}).  The decoupling of the gap and the tilt term leaves crucial signatures on the optically activated momentum surface. As a result, CPGE trace profile with  frequency changes for these models such as CPGE trace acquires finite value at smaller frequency with increasing tilt for
model II as compared to model I when chemical potential is set at halfway between the WNs. 
Apart from the frequency selection rule, as dictated by $ \delta(\hbar\omega - E_{\bm k,12})$, the remaining factor $\Delta f_{\bm k,12}  \Delta v_{{\bm k},12}^i\Omega_{{\bm k},i}$ determines the magnitude of CPGE. The specific structure of $\sigma_0$ term  imprints its effect
in the frequency profile of CPGE via $\Delta f_{\bm k,12}$. We find that 
in model II,  the magnitude of the CPGE is higher as compared to that for model I as far as the type-II phase is concerned.

On the other hand, as restricted  by the mirror symmetry constraints,  
the diagonal components of BCD mediated second order responses only remains non-zero. The  peak and dip structures of the BCD, appearing around the WNs, are   
directly related the  chiralities of the  WNs (see Figs.~\ref{fig:bcd_modelI} and \ref{fig:bcd_modelII}).
This is similar to the CPGE trace where the sign of the quantized plateau is determined by the 
chirality of the activated WN. The gap term being coupled (decoupled) with the tilt term leads to qualitatively similar (different) response between different BCD components.   
The transport coefficients acquire higher values
with enhancing the non-linearity in dispersion. Moreover, the significantly different  BCD response  between model I and  II can be traced back to their 
distinct BCD density profile resulted from the non-identical  velocity factors.

Having investigated the diffusive transport, we  explore the ballistic transport MHE in the later part of our work. We find that the Fermi surface contribution is significantly modified whether the gap and tilt term are mutually coupled or decoupled. The signature of the gap term is very clearly manifested in the MHC for model II whereas in model I, the MHC profile exhibits complicated structures (see Figs.~\ref{fig:mhc_modelI} and \ref{fig:mhc_modelII}).  The MHC for higher topological charge is not found to be always larger than that for the lower topological charge. The distribution of BC for the selected momentum modes over the BZ dictates the chemical potential dependence of the MHC. Therefore,  the momentum modes away from the WNs,  combined with the Fermi surface characteristics, play an intriguing role to determine the MHC. The effects of the WNs are  very pronounced in the CPGE and BCD while  the MHE can be significantly stronger away from the  WN's energies. 

\section{Acknowledgements}
TN thanks Snehasish Nandy for discussions.  
 DMK acknowledges the Deutsche Forschungsgemeinschaft (DFG, German Research Foundation) for support through RTG 1995, within the Priority Program SPP 2244 ``2DMP'' and under Germany's Excellence Strategy - Cluster of Excellence Matter and Light for Quantum Computing (ML4Q) EXC 2004/1 - 390534769.

\bibliography{ref}

\end{document}